\documentclass[twocolumn, twocolappendix, numberedappendix]{aastex63}

\newcommand\submitms{y} 
\if\submitms y
\else
\usepackage{apjfonts}
\fi

\usepackage{ifthen}
\usepackage{afterpage}

\newcommand{\comment}[1]{}

\usepackage{amsmath}
\usepackage{natbib}
\usepackage{appendix}
\usepackage{gensymb}

\providecommand{\adsurl}[1]{\href{#1}{ADS}}

\DeclareSymbolFont{UPM}{U}{eur}{m}{n}
\DeclareMathSymbol{\umu}{0}{UPM}{"16}
\let\oldumu=\umu
\renewcommand\umu{\ifmmode\oldumu\else$\oldumu$\fi}
\newcommand\micro{\umu}
\renewcommand\micron{\micro m}
\newcommand\microns{\micro m}

\renewcommand{\color}[1]{}

\if\submitms y
\else
\comment{\hfill\textbf{DRAFT of {\today} \now}.}
\fi

\shorttitle{Elliptical Photometry with \textit{Spitzer}}
\shortauthors{Challener {\em et al.}}

\begin{document}

\title{Identification and Mitigation of a Vibrational Telescope Systematic with Application to Spitzer}
\author[0000-0002-8211-6538]{Ryan C. Challener}
\affiliation{Planetary Sciences Group, Department of Physics, 
  University of Central Florida, Orlando, FL 32816-2385}

\author[0000-0002-8955-8531]{Joseph Harrington}
\affiliation{Planetary Sciences Group, Department of Physics, 
  University of Central Florida, Orlando, FL 32816-2385}

\author[0000-0003-2733-8725]{James Jenkins}
\affiliation{Departamento de Astronoma, Universidad de Chile,
  Camino El Observatorio 1515, Las Condes, Santiago, Chile}
\affiliation{Centro de Astrof\'isica y Tecnolog\'ias Afines (CATA),
  Casilla 36-D, Santiago, Chile}

\author{Nicol\'as T. Kurtovic}
\affiliation{Departamento de Astronoma, Universidad de Chile,
  Camino El Observatorio 1515, Las Condes, Santiago, Chile}

\author{Ricardo Ramirez}
\affiliation{Departamento de Astronoma, Universidad de Chile,
  Camino El Observatorio 1515, Las Condes, Santiago, Chile}

\author[0000-0002-8307-144X]{Kathleen J. McIntyre}
\affiliation{Planetary Sciences Group, Department of Physics, 
  University of Central Florida, Orlando, FL 32816-2385}

\author[0000-0002-9338-8600]{Michael D. Himes}
\affiliation{Planetary Sciences Group, Department of Physics, 
  University of Central Florida, Orlando, FL 32816-2385}

\author[0000-0001-6827-9077]{Eloy Rodr\'iguez}
\affiliation{Instituto de Astrof\'isica de Andaluc\'ia (IAA, CSIC)
  Glrorieta de la Astronom\'ia, s/n E-18008 Granada, Spain}

\author{Guillem Anglada-Escud\'e}
\affiliation{School of Physics and Astronomy, Queen Mary University
  of London, 327 Mile End Road, London E1 4NS, UK}

\author{Stefan Dreizler}
\affiliation{Institut f\"ur Astrophysik, Georg-August-Universit\"at
  G\"ottingen Friedrich-Hund-Platz 1, 37077 G\"ottingen, Germany}

\author[0000-0002-9152-5042]{Aviv Ofir}
\affiliation{Department of Earth and Planetary Sciences, Weizmann
  Institue of Science, 234 Herzl Street, Rehovot 76100, Israel}

\author{Pablo A. Pe\~na Rojas}
\affiliation{Departamento de Astronoma, Universidad de Chile,
  Camino El Observatorio 1515, Las Condes, Santiago, Chile}

\author[0000-00002-6689-0312]{Ignasi Ribas}
\affiliation{Institut de Ci\'encies de l'Espai (ICE, CSIC),
  C/Can Magrans s/n, Campus UAB, 08193 Bellaterra, Spain}
\affiliation{Institut d'Estudis Espacials de Catalunya (IEEC), 08034
  Barcelona, Spain}

\author[0000-0002-1607-6443]{Patricio Rojo}
\affiliation{Departamento de Astronoma, Universidad de Chile,
  Camino El Observatorio 1515, Las Condes, Santiago, Chile}

\author[0000-0002-4365-7366]{David Kipping}
\affiliation{Department of Astronomy, Columbia University, 550 W
  120th Street, New York, NY 10027}

\author{R. Paul Butler}
\affiliation{Earth \& Planets Laboratory, Carnegie Institution
  for Science, 5241 Broad Branch Road NW, Washington D.C. 
  20015-1305, USA}

\author{Pedro J. Amado}
\affiliation{Instituto de Astrof\'isica de Andaluc\'ia (IAA, CSIC)
  Glrorieta de la Astronom\'ia, s/n E-18008 Granada, Spain}

\author{Cristina Rodr\'iguez-L\'opez}
\affiliation{Instituto de Astrof\'isica de Andaluc\'ia (IAA, CSIC)
  Glrorieta de la Astronom\'ia, s/n E-18008 Granada, Spain}

\author{Enric Palle}
\affiliation{Instituto de Astrof\'isica de Canarias (IAC), E-38205
  La Laguna, Tenerife, Spain}
\affiliation{Departamenta de Astrof\'isica, Universidad de La Laguna
  (ULL), E-38206 La Laguna, Tenerife, Spain}

\author{Felipe Murgas}
\affiliation{Instituto de Astrof\'isica de Canarias (IAC), E-38205
  La Laguna, Tenerife, Spain}
\affiliation{Departamenta de Astrof\'isica, Universidad de La Laguna
  (ULL), E-38206 La Laguna, Tenerife, Spain}


\begin{abstract}

  We observed Proxima Centauri with the \textit{Spitzer Space
    Telescope} InfraRed Array Camera (IRAC) five times in 2016 and
  2017 to search for transits of Proxima Centauri b. Following
  standard analysis procedures, we found three asymmetric,
  transit-like events that are now understood to be vibrational
  systematics. This systematic is correlated with the width of the
  point-response function (PRF), which we measure with rotated and
  non-rotated Gaussian fits with respect to the detecor array. We show
  that the systematic can be removed with a novel application of an
  adaptive elliptical-aperture photometry technique, and compare the
  performance of this technique with fixed and variable
  circular-aperture photometry, using both BiLinearly Interpolated
  Subpixel Sensitivity (BLISS) maps and non-binned Pixel-Level
  Decorrelation (PLD). With BLISS maps, elliptical photometry results
  in a lower standard deviation of normalized residuals, and reduced
  or similar correlated noise when compared to circular apertures. PLD
  prefers variable, circular apertures, but generally results in more
  correlated noise than BLISS. This vibrational effect is likely
  present in other telescopes and \textit{Spitzer} observations, where
  correction could improve results. Our elliptical apertures can be
  applied to any photometry observations, and may be even more
  effective when applied to more circular PRFs than
  \textit{Spitzer's}.
\end{abstract}


\section{INTRODUCTION}
\label{sec:introduction}


Exoplanet science has pushed the \textit{Spitzer Space Telescope}
\citep{WernerEtal2004apjsSpitzer} far beyond its initial
design. Transiting and eclipsing exoplanet signals are on the order of
1\% and 0.1\% of their host star, respectively, far below expected
performance of the InfraRed Array Camera (IRAC,
\citealp{FazioEtal2004apjsIRAC}). Soon, the \textit{James Webb Space
  Telescope} (JWST) will provide an unprecedented combination of
spectral resolution, {\color{red} wavelength coverage}, collecting
area, and stability for exoplanet science
\citep{DemingSeager2017JGREreview}. {\color{red} Currently, the field
  is limited to 1D characterization of most planets
  \citep[e.g.,][]{KreidbergEtal2018apjWASP107b}, 2D thermal mapping of
  very hot targets \citep{DeWitEtal2012aaHD189Map,
    MajeauEtal2012apjlHD189Map}, and 3D characterization with
  \textit{Hubble Space Telescope} data of the hottest known planets
  \citep[e.g.,][]{StevensonEtal2014sciWASP43bphasecurve,
    StevensonEtal2017ajWASP43bPhaseCurve}. JWST will significantly
  improve planetary characterization possiblities, but} small and cold
planets will still be a challenge. Hotter terrestrial targets, like
TRAPPIST-1b \citep{GillonEtal2016natTRAPPIST,
  GillonEtal2017natTRAPPIST}, will require ${\sim}10$ secondary
eclipses for a confident detection \citep{MorleyEtal2017apjJWSTearth},
but temperate Earth-like targets will be difficult, if not impossible
\citep{RauerEtal2011A&Abiosig, RugheimerEtal2015ApJuvprints,
  BatalhaEtal2018ApJjwsttemp, Beichman&Greene2018haexJWST}. We must
take advantage of every technique available if we hope to characterize
these planets.

\textit{Spitzer} IRAC suffers from two primary systematic effects: an
easily-removed ``ramp'' that causes measured flux to vary with time,
and an intrapixel gain variation that creates correlations between
flux and target position on the detector at a subpixel level
\citep[e.g.,][]{CharbonneauEtal2005apjTrES1}. These effects are
present in all 3.6 and 4.5 \micron\ observations, although the ramp is
sometimes weak enough to be ignored. Several independent modeling
techniques, such as {\color{red} Gaussian-weighted maps
  \citep{BallardEtal2010paspGJ436c, LewisEtal2013apjHATP2b},}
BiLinearly Interpolated Subpixel Sensitivity (BLISS,
\citealp{StevensonEtal2012apjBLISShd149b}), Pixel-Level Decorrelation
(PLD, \citealp{DemingEtal2015apjHATP20pld}), and Independent Component
Analysis \citep{Morello2015apjICA} have successfully removed the
position-correlated noise, enabling transiting exoplanet observations
with uncertainties $<$100 ppm and retrieving accurate planetary
parameters \citep{IngallsEtal2016ajSpitzerSystematics}.

A third, much less common effect creates light-curve features that
resemble transiting exoplanets. This effect has been linked to
activity in the ``noise pixel'' parameter \citep{Mighell2005mnrasPSFs,
  LewisEtal2013apjHATP2b}, a measurement of the number of pixels that
contribute to centering and photometry, or the size of the
point-response function (PRF). Spikes in the noise pixels are known to
correlate with transit-like signals, and are likely caused by
high-frequency telescope oscillations of unknown origin (see
irachpp.spitzer.caltech.edu/page/np\_spikes). {\color{red} Previous
  studies show that PRF Gaussian-width dependent models can account
  for time-evolving point-source shapes
  \citep[e.g.,][]{LanotteEtal2014aapGJ436b, MendoncaEtal2018ajWASP43b,
    MansfieldEtal2020apjKELT9b}, which may be related to the
  noise-pixel effect.}

\begin{figure*}[t]
  \includegraphics[width=7.5in]{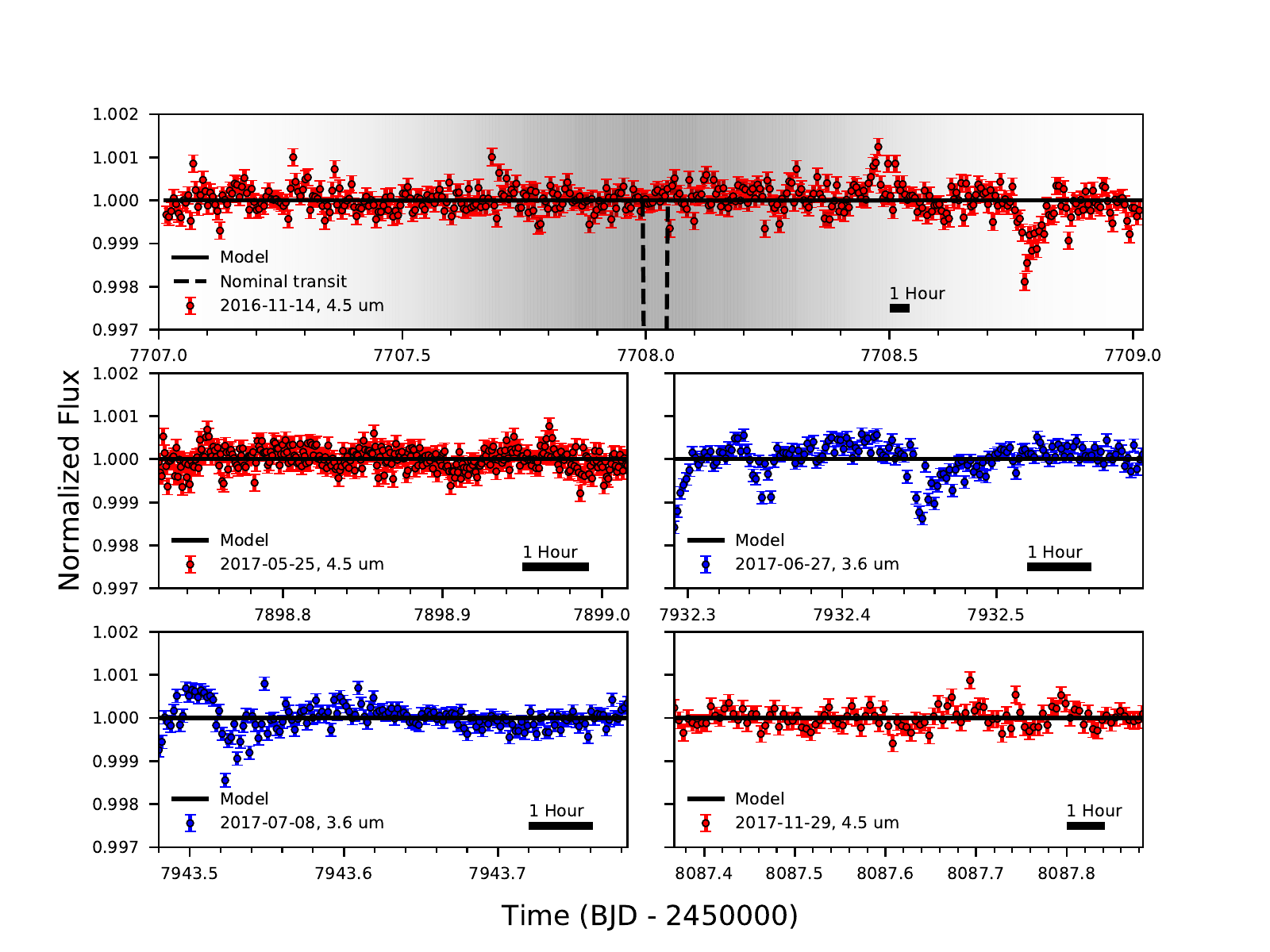}
  \caption{The five \textit{Spitzer} IRAC observations of Proxima,
    using fixed circular apertures and a BLISS map, binned to 1500
    frames per data point. {\color{red} From top left to bottom right,
      aperture radii are 3.00, 2.25, 3.25, 4.50, and 2.75 pixels.} We
    divided out the BLISS map and ramp model where appropriate, so
    ideally the resulting light curve should be flat (matching the
    black model line). Observation dates and channels appear on each
    plot. The dashed line in the top panel marks the nominal Proxima b
    transit and the shaded region denotes the uncertainty on transit
    time (at the time of observation).}
  \label{fig:subplots}
\end{figure*}

We observed Proxima Centauri (hereafter Proxima) in 2016 and 2017 with
\textit{Spitzer} IRAC to search for transits of the planet Proxima b
\citep{Anglada-EscudeEtal2016NatProxb}.
\cite{JenkinsEtal2019mnrasProxNoTran}, hereafter J19, presented the
results from the first observation. When following standard data
reduction procedures {\color{red} and dividing out best-fitting
  systematic models (with only a flat astrophysical model)}, these
observations contain three transit-like events (see Figure
\ref{fig:subplots}) that resemble the asymmetric shapes created by
transits of evaporating planets
\citep[e.g.,][]{RappaportEtal2014apjKOI2700b,
  SanchisOjedaEtal2015apjK2-22b, VanderburgEtal2015natDisintWD} or
comets \citep[e.g.,][]{RappaportEtal2018mnrasExocomets}. We now know,
conclusively, that these events are localized systematic effects due
to high-frequency telescopic vibration of unknown origin. When the
telescope vibrates, the PRF smears along the direction of the
vibration. During the vibration, fixed-radius photometry apertures
spill light, resulting in lower measured flux with larger vibrational
amplitudes.

In this work, we present evidence that the systematic is due to
vibration, several new methods to identify when this vibrational
systematic occurs, and a new aperture photometry method to correct
it. The paper is organized as follows: in Section \ref{sec:obs} we
describe our observations, in Section \ref{sec:diag} we discuss how to
identify this systematic, in Section \ref{sec:treat} we present our
elliptical photometry method, in Section \ref{sec:disc} we interpret
our findings, and in Section \ref{sec:results} we lay out our
conclusions.

\section{OBSERVATIONS}
\label{sec:obs}

\begin{table*}[ht]
\movetableright=0.3in
\caption{\label{tbl:obs}Observations}
\begin{tabular}{lccccc}
\hline\hline
                           & Nov. 2016    & May 2017     & June 2017    & July 2017    & Nov. 2017 \\
\hline
Wavelength (\microns\ )    & $4.5$        & $4.5$        & $3.6$        & $3.6$        & $4.5$     \\
Obs. Start (MBJD\tablenotemark{a}) & $7707.01325$ & $7898.72171$ & $7932.29024$ & $7943.47907$ & $8087.38757$\\
Obs. Duration (hours)      & $48.04$      & $7.07$       & $7.34$       & $7.34$       & $12.52$   \\
Frame Time (s)             & 0.02         & 0.02         & 0.02         & 0.02         & 0.02 \\
\hline
\end{tabular}
\tablenotetext{a}{MBJD = Modified Barycentric Julian Date = BJD - 2450000.}
\end{table*}

We observed Proxima with the \textit{Spitzer} InfraRed Array Camera,
with both the 3.6 and 4.5 \microns\ filters, for a total of $>80$
hours (Table \ref{tbl:obs}). The 48-hour stare bracketed the predicted
transit time of Proxima b, and shorter observations occurred at times
when further transits should occur, if the feature in the stare was
caused by Proxima b \citep{Anglada-EscudeEtal2016NatProxb}. All
observations were in $32 \times 32$ subarray mode and centered on the
IRAC ``sweet spot'', at (-0.352\arcsec, 0.064\arcsec) and
(-0.511\arcsec, 0.039\arcsec) for 3.6 \microns\ and 4.5 \microns,
respectively. We bracketed each science observation with an initial
30-minute observation to minimize telescope pointing settling and a
final 10-minute observation, for those who wish to generate their own
dark frames, per \textit{Spitzer} Science Center recommendations.

Notably, due to the brightness of the target, our observations
utilized the shortest frame time, 0.02 seconds, which allows temporal
resolution of high-frequency effects. To handle the large data volume
from this cadence, our observations have data gaps. The 48 hour 4.5
\microns\ stare has 17-second gaps between 64-frame subarray chunks,
24-second gaps between Astronomical Observation Requests (AORs), and
${\sim}4$ minute gaps every 16 hours for data downlink and target
reacquisition. Both 3.6 \microns\ observations have 6 second gaps
between subarray chunks and 14 second gaps between AORs. The shortest
4.5 \microns\ observation has 2 -- 2.5-second gaps between subarray
chunks, and only one AOR. The November 2017 {\color{red} 4.5
  \micron\ } observation has the same gaps as the 48-hour stare,
without the downlink and target reacquisition.

The telescope's heater, which introduces motion on the detector in
{$\sim$}40-minute cycles, was turned off for the duration of all five
observations, following then-current \textit{Spitzer} procedures for
exoplanet observations.  This minimizes the impact of the intrapixel
systematic, allowing closer study of other effects.

\section{CENTERING AND PHOTOMETRY}
\label{sec:centphot}

We use {\color{red} the} Photometry for Orbits, Eclipses,
and Transits code (POET,
\citealp[e.g.,][]{StevensonEtal2012apjBLISShd149b,
  BlecicEtal2013apjWASP14b, CubillosEtal2013ApjWASP8b,
  BlecicEtal2014apjWASP43b, CubillosEtal2014apjTrES1b,
  HardyEtal2017apjHATP13b}) for all analyses herein.  The steps in
producing light curves are bad pixel identification, determining the
location of the target (centering), and measuring its brightness
(photometry). {\color{red} We identify bad pixels by performing a
  twice-iterated 4$\sigma$ rejection on 64-frame chunks, and discard
  these pixels from further analysis.} This work relies heavily on
different centering and photometry methods, so we describe our
implementations in detail.

\subsection{CENTERING METHODS}
\label{sec:cent}

We apply four centering methods: Gaussian fitting, rotated-Gaussian
fitting, center-of-light, and least asymmetry
\citep{LustEtal2014paspLeastAsym}.  Our Gaussian fitting includes
parameters for $x$ and $y$ position, widths in both dimensions, a
height, and a constant background level. Center-of-light calculates an
average position, weighted by the brightness of each pixel, much like
a center-of-mass calculation. Least asymmetry computes an asymmetry
value for each pixel by considering the symmetry of surrounding flux
values and then fitting an inverted Gaussian to determine the point of
least asymmetry. Our rotated-Gaussian fitting is described further in
Section \ref{sec:diag}. Unless stated otherwise, we perform centering
on a 17$\times$17 pixel box around the target. Least asymmetry uses a
9$\times$9 pixel box to calculate the asymmetry of a given pixel in
the 17$\times$17 centering box. {\color{red} We did not vary these box
  sizes in this analysis, and they are consistent with previous works
  \citep[e.g.,][]{CubillosEtal2014apjTrES1b}. Frames with significant
  positional outliers and frames where a good centering fit could not
  be achieved were removed from the analysis prior to modeling.}

\subsection{PHOTOMETRY METHODS}
\label{sec:phot}

Since the IRAC point-spread function (PSF) is undersampled, we
bilinearly interpolate all masks and images to 5$\times$ resolution,
ensuring that flux is conserved. For the Boolean masks {\color{red}
  (the IRAC-supplied bad pixel masks combined with our flagged bad
  pixels)}, any interpolated value less than one is considered a bad
subpixel (i.e., any subpixel between the center of a bad pixel and a
good pixel are marked as bad). We then perform aperture photometry on
the interpolated, masked images, increasing all relevant length scales
by the interpolation factor (aperture radius, background annulus
radii, etc.), so the apertures include subpixels. We calculate the
background level as a mean of the pixels in a 7 -- 15-pixel annulus
around the centering position, {\color{red} consistent with previous
  analyses \citep[e.g.][]{BlecicEtal2013apjWASP14b,
    CubillosEtal2013ApjWASP8b, HardyEtal2017apjHATP13b}}.

We use three aperture photometry methods: fixed, variable
\citep{LewisEtal2013apjHATP2b}, and elliptical (see Section
\ref{sec:treat}). Fixed photometry uses a constant-size aperture
throughout a given observation. We use fixed aperture radii from 1.5
to 4.5 pixels, in steps of 0.25 pixels. Variable photometry derives
aperture radii from the same 17$\times$17 box used for centering, as
described by \cite{LewisEtal2013apjHATP2b}. Our variable-aperture
radii are calculated as

\begin{equation}
  \label{eqn:varrad}
  R_{\rm var} = a\sqrt{\beta}+b
\end{equation}

\noindent
where $a$ is a scaling factor from 0.5 -- 1.5 in steps of 0.25, $b$ is
an offset from -1.0 -- 2.0 pixels in steps of 0.5, $\beta$ is the
noise-pixel parameter {\color{red} \citep{Mighell2005mnrasPSFs}}, defined as

\begin{equation}
  \label{eqn:noispix}
  \beta = \frac{\sum_i \left(I(i)\right)^2}{\left(\sum_i I(i)^2\right)},
\end{equation}

\noindent
where $I(i)$ is the intensity at pixel $i$, considering all pixels
within the centering aperture. $\sqrt{\beta}$ is $\sim$2 pixels on average
and varies by $\sim$0.2 pixels throughout an observation. Calculation
of $\beta$ should be done after background subtraction, as this
significantly reduces scatter in the aperture radii and noise in the
light curve.

Elliptical apertures vary in size similarly to variable apertures, but
use the 1$\sigma$ widths of the Gaussian centering fit as the base,
rather than $\sqrt{\beta}$. Then, the elliptical apertures are calculated
as

\begin{align}
  R_{{\rm ell},x} = a\sigma_x+b \nonumber \\
  R_{{\rm ell},y} = a\sigma_y+b \label{eqn:ellrad}
\end{align}

\noindent
where $\sigma_x$ and $\sigma_y$ are the ellipse widths in $x$ and $y$
(which vary in time), $a$ ranges from 3 -- 7 in steps of 1, and $b$
again covers -1.0 -- 2.0 pixels in steps of 0.5 pixels. The ellipse
widths typically range from 0.5 -- 0.6 pixels during an observation.
See Section \ref{sec:treat} for a more in-depth description of
elliptical photometry.

{\color{red} Regardless of photometry method, we discard any frame
  which contains bad pixels within the aperture. We do not discard any
  additional frames due to flux variation.}

We use small apertures to avoid additional noise from background
pixels, but they necessitate an aperture correction to account for
lost light. With fixed apertures, we rescale the final photometry
based on the fraction of the interpolated PSF in the aperture. For
variable and elliptical photometry, we rescale on the same principle,
using an average aperture size and shape. It is possible to rescale
the photometry using time-variable apertures, but this negates the
correction made by the photometry methods. The interpolated PSF,
provided by the \textit{Spitzer} Science Center, is constant, but we
suspect the true PSF stretches on short timescales, making accurate
rescaling on a frame-by-frame basis impossible (see Section
\ref{sec:treat} for further discussion). Regardless, we are interested
in the relative photometry, not the absolute, so whether or not we
scale by a constant factor has little bearing on this work.

We choose the best centering and photometry methods by minimizing the
binned-$\sigma$ $\chi^2$ (hereafter $\chi^2_{\mathrm{bin}}$;
\citealp{DemingEtal2015apjHATP20pld}) of a best-fit light-curve
model. In brief, this metric searches for model residuals that behave
like white noise. White noise, measured by the standard deviation of
normalized residuals (SDNR), predictably scales as $1/\sqrt{n}$, where
$n$ is the number of items in each bin.  {\color{red} Therefore, we
  calculate log(SDNR) vs.\ log($n$) for a range of $n$, then calculate
  the $chi^2$ of a line with a slope of $-1/2$, anchored to the
  log(SDNR) of the unbinned residuals ($n=1$).} We repeat this process
for every light curve produced by each unique combination of centering
and photometry methods, and take the best fit (lowest
$\chi^2_{\mathrm{bin}}$) as optimal. See
\citealp{DemingEtal2015apjHATP20pld} and Appendix \ref{app:bsig} for a
complete description of the calculation.

{\color{red} We also tested a calculation of average relative
  correlated noise as a metric for photometry extraction method
  optimization. For this metric, we calculate the expected white noise
  residual RMS \citep{PontEtal2006mnrasRedNoise, WinnEtal2008apjXO3b}
  and measured residual RMS over all bin sizes from 1 frame to half
  the length of the observation, and compute the average factor
  between the two (see Figure \ref{fig:rmsexample} for an
  example). With BLISS models, this metric results in the same
  photometry methods with identical or larger aperture sizes compared
  to $\chi^2_{\textrm{bin}}$. Limiting the bin sizes to ten minutes or
  greater (time scales roughly relevant to eclipse, transit, and phase
  curve observations) did not change the result.}

{\color{red} In addition, we experimented with a metric similar to
  $\chi^2_{\textrm{bin}}$ but with the theoretical $1/\sqrt{n}$ line
  anchored to the expected white noise unbinned residual RMS rather
  than the measured unbinned SDNR. Thus, this metric searches for
  photometric extractions that behave like white noise at all bin
  sizes, rather than those which only must behave similarly at large bin
  sizes. With BLISS, this metric selects identical extractions as
  $\chi^2_{\textrm{bin}}$ (both method and aperture size). Since both
  these metrics led to similar results, we present only the results
  using $\chi^2_{\textrm{bin}}$.}

\section{LIGHT-CURVE MODELING}

We modeled our light curves with both PLD and BLISS to correct the
intrapixel systematic and to assess each model's ability to address
the vibrational systematic. BLISS maps correct for intrapixel
sensitivity variations by gridding the detector into fine subpixels.
We assign each frame to a subpixel based on the target position from
centering, compute the sensitivity of each subpixel based on the
average flux of all frames associated with them, once all other models
(astrophysical or otherwise) have been removed, and bilinearly
interpolate the sensitivity grid to find a correction factor for each
frame. The generic full model formula is

\begin{equation}
  \label{eq:bliss}
  F(x,y,t) = F_s A(t) M(x,y) R(t),
\end{equation}

\noindent
where $x$ and $y$ are the target positions in each frame, $t$ is time,
$F_s$ is stellar flux, $A$ is the astrophysical model, $M$ is the
BLISS map, and $R$ is the time-dependent ramp. In a typical transiting
exoplanet analysis, $A(t)$ would be a combination of transits,
eclipses, and phase curve variation models, but in this case, there
are no modeled astrophysical variations.

PLD removes the same effect by treating the data as a weighted sum of
normalized pixel values, where the weights are free parameters of the
model. The model is

\begin{equation}
  F(t) = F_s\left(\sum_i^{n_p} c_j \hat P_j  + A(t) + R(t)\right),
\end{equation}

\noindent
where $n_p$ is the number of pixels considered, $c_j$ are pixel
weights, and $\hat P_j$ are time-dependent normalized pixel
values. {\color{red} In this work, we use the 9 brightest pixels ($n_p
  = 9$) in our PLD models. Although it is common to bin the light
  curves temporally when using PLD
  \citep[e.g.][]{DemingEtal2015apjHATP20pld, WongEtal2015apjWASP14},
  we do not bin the data in order to separate the models' ability to
  remove correlated noise from the effects of binning (see Section
  \ref{sec:disc} for further discussion).} See
\cite{StevensonEtal2012apjBLISShd149b} and
\cite{DemingEtal2015apjHATP20pld} for in-depth descriptions of BLISS
and PLD, respectively.

Figure \ref{fig:subplots} shows the fixed-aperture light curves,
modeled with BLISS, to highlight the vibrational
systematic. {\color{red} Aperture radii are 3.00 pixels (November 2016
  4.5 \microns), 2.25 pixels (May 2017 4.5 \microns), 3.25 pixels
  (June 2017 3.6 \microns), 4.50 pixels (July 2017 4.5 \microns), and
  2.75 pixels (November 2017 4.5 \microns).} The systematic is present
in the November 2016 {\color{red} 4.5 \micron}, June 2017
{\color{red} 3.6 \micron}, and July 2017 {\color{red} 3.6 \micron}
light curves, so we focus on these observations going
forward. {\color{red} While the vibration appears to be more common at
  3.6 \microns\ (2 -- 3 occurrences) than at 4.5 \microns\ (1
  occurrence), possibly due to a narrower PRF, our sample size is only
  large enough to confirm that the vibration is not limited to one
  bandpass.}

\section{SYSTEMATIC DIAGNOSTICS}
\label{sec:diag}

Past works used the noise pixel measurement (Equation
\ref{eqn:noispix}) to identify activity in the PRF, and correct for it
with variable-aperture photometry
\citep[e.g.,][]{LewisEtal2013apjHATP2b, DemingEtal2015apjHATP20pld,
  GarhartEtal2018aapQatar1b,
  JenkinsEtal2019mnrasProxNoTran}. Effectively, noise pixels measure
the number of pixels above the background (contributing to centering
and photometry). A wider (narrower) PRF should result in a larger
(smaller) noise-pixel value. Since noise pixels measure an area, the
radius of the photometry aperture required for the PRF is the root of
the noise pixels, commonly with additional multiplicative and/or
additive scaling (see Section \ref{sec:phot}). Thus, as the PRF size
varies throughout the observation, so does the photometry aperture
radius.

\begin{figure}
  \includegraphics[width=3.25in]{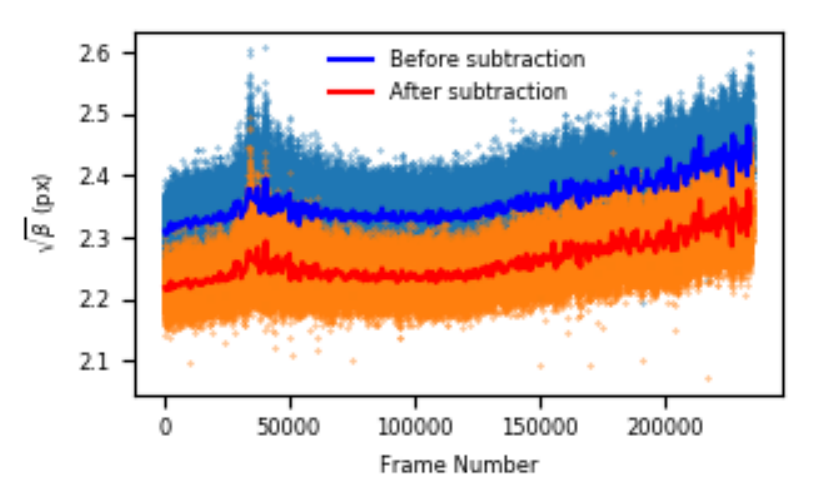}
  \caption{\color{red} A comparison of $\sqrt{\beta}$ in the July 3.6
    \microns\ observation, when calculated before and after background
    subtraction. Solid lines are binned to 500 points. Scatter,
    measured by the standard deviation, decreases by $\sim 14$\% when
    background subtraction is done before calculation of $\beta$.}
  \label{fig:npbgsub}
\end{figure}

J19 found that, using common techniques, centering and photometry
selection criteria selected against variable photometry apertures. We
have since improved the variable-aperture photometry by calculating
the aperture radii after background subtraction, which reduces
uncertainty introduced by unimportant pixels. {\color{red} Figure
  \ref{fig:npbgsub} shows a comparison of aperture radii
  ($\sqrt{\beta}$ in Equation \ref{eqn:varrad}) in the July 3.6
  \micron\ observation when $\beta$ is computed before and after
  background subtraction. In this case, the standard deviation of
  $\sqrt{\beta}$ decreases by $\sim 14$\%.} With this improvement,
variable-aperture radii are preferred over fixed-aperture radii
{\color{red} for these observations}, although they still introduce
noise to the light curve due to the additional noise-pixel parameter.
{\color{red} It is unclear if \cite{LewisEtal2013apjHATP2b}, who
  introduced variable photometry apertures, calculated noise pixels
  and aperture radii before or after background subtraction.}

\begin{figure*}
  \includegraphics[width=0.33\textwidth]{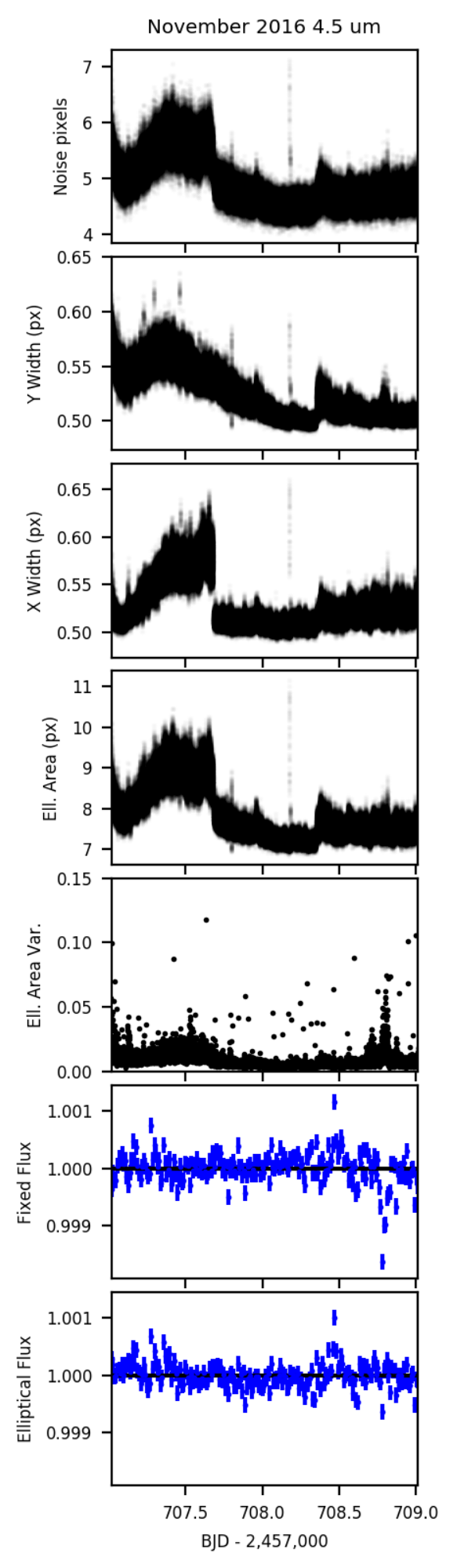}
  \includegraphics[width=0.33\textwidth]{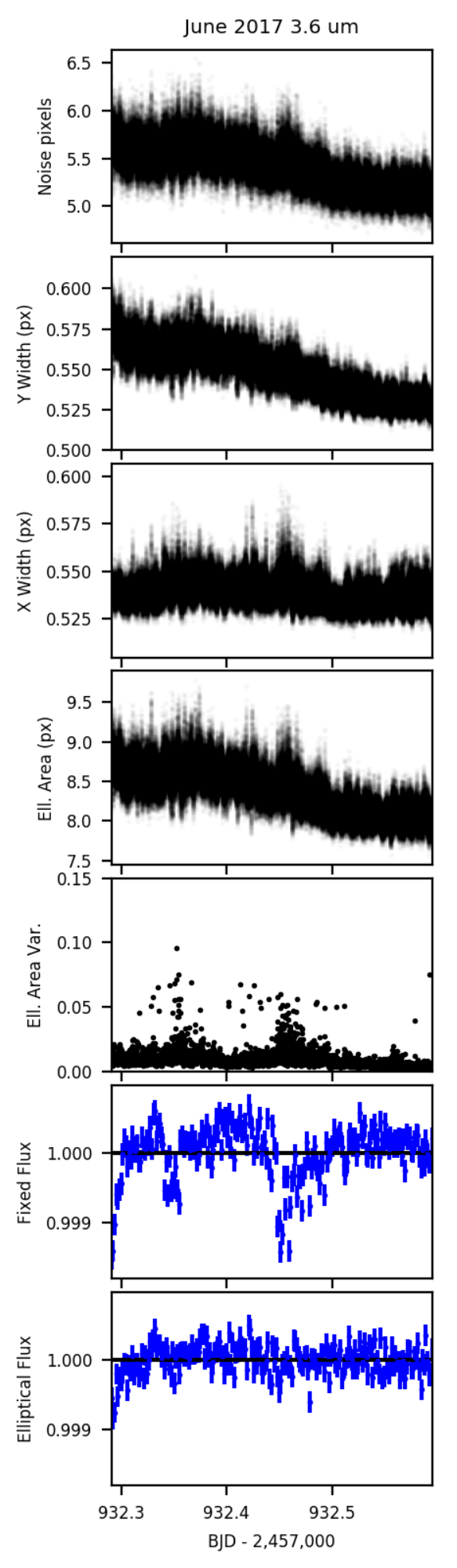}
  \includegraphics[width=0.33\textwidth]{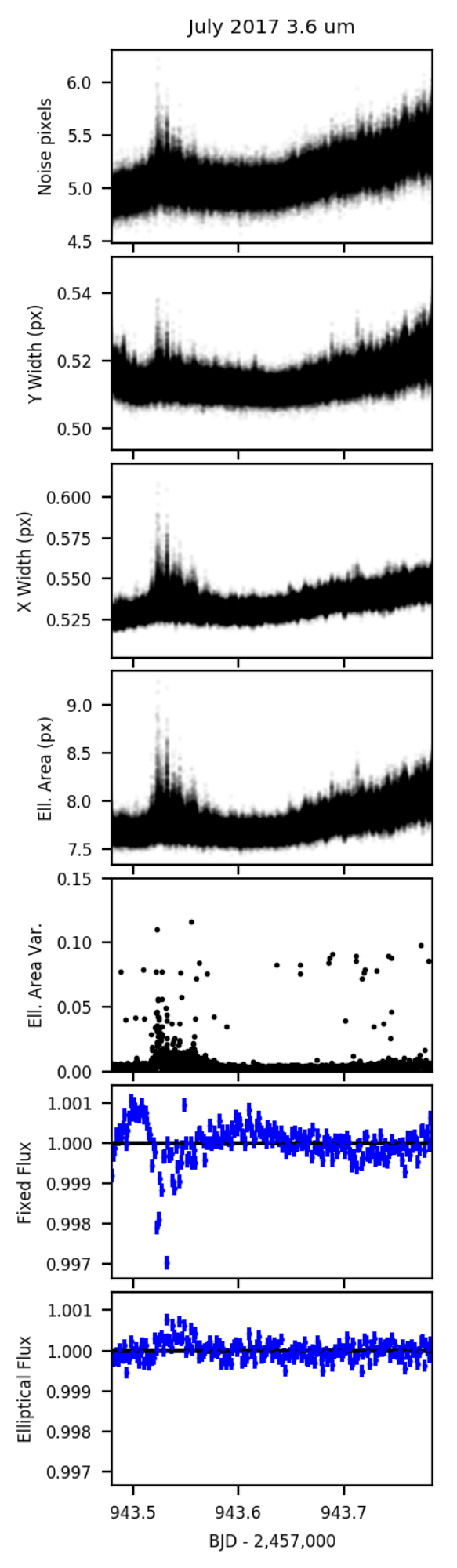}
  \caption{Systematic identification methods for the three
    observations that contain the systematic. Columns from left to
    right are the November 2016 {\color{red} 4.5 \micron}, June 2017
    {\color{red} 3.6 \micron}, and July 2017 {\color{red} 3.6
      \micron} observations. From top to bottom, rows are noise
    pixels, PRF $y$ width, PRF $x$ width, elliptical area, elliptical
    area variance, the best fixed-aperture light curve, and the best
    elliptical-aperture light curve.}
  \label{fig:identmeth}
\end{figure*}

Oscillations in the telescope, if higher frequency than the exposure
time, could be hidden from centering, but they would be evident in a
widening of the PRF in the direction of the vibration. By fitting a
Gaussian to the PRF, we determine 1$\sigma$ widths in $x$ and $y$ (see
Figure \ref{fig:identmeth}, second and third rows) and notice a
prominent widening in the PRF at the time of the systematic. This
widening is even more evident in a measure of the 3$\sigma$ area of
the Gaussian, which we compute as an ellipse with axes along the $x$
and $y$ directions (see Figure \ref{fig:identmeth}, fourth row). We
also measure the variance in this elliptical area, on a 64-frame
basis, to look for PRF activity (see Figure \ref{fig:identmeth}, fifth
row). 

\begin{figure} 
  \includegraphics[width=3.25in]{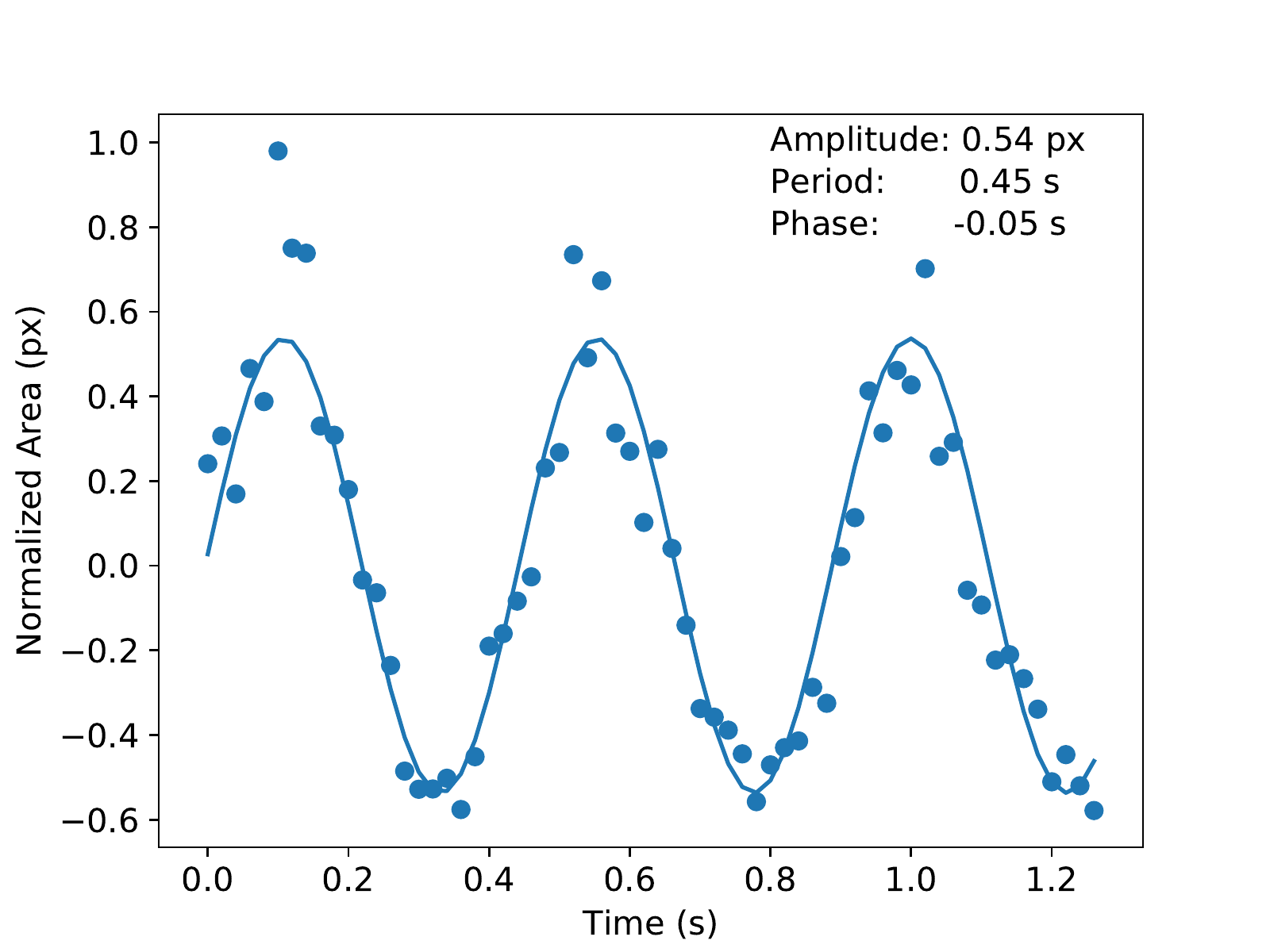}
  \caption{Mean-subtracted Gaussian elliptical area of a single chunk
    of 64 frames during the peak of the systematic in the July 2017
    {\color{red} 3.6 \micron} observation. We fit a simple sinusoid
    and determined a 0.45 second periodicity.}
  \label{fig:sinefit}
\end{figure}

Our short exposures (0.02 seconds) allow temporal resolution of
high-frequency effects. Figure \ref{fig:sinefit} shows the 3$\sigma$
elliptical area of a single set of 64 frames during the peak of the
systematic in the July 2017 {\color{red} 3.6 \micron}
observation. We find a clear sinusoidal pattern with a period of 0.45
seconds, evidence for telescope oscillation. Since the shape of the
PRF is changing, and the photometric effect is a net loss in flux,
integrating exposures by a multiple of the oscillation timescale will
not correct the effect.

\begin{figure*}
  \includegraphics[width=7.5in]{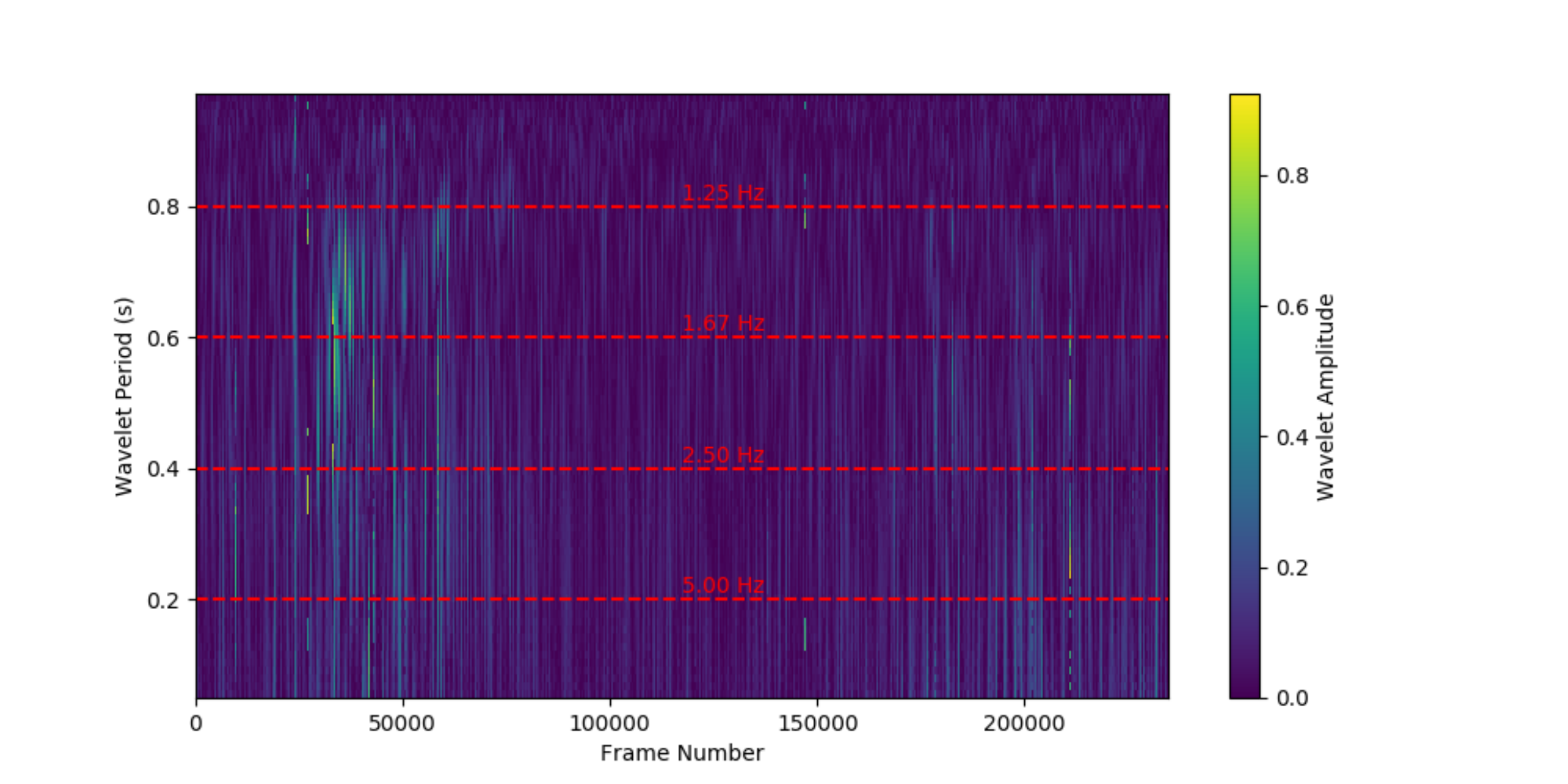}
  \caption{Continuous Morlet wavelet transform of the July 2017
    {\color{red} 3.6 \micron} Proxima observation Gaussian elliptical
    area with a fixed 2.5 pixel photometry aperture radius and
    non-rotated Gaussian centering. The activity {\color{red} (higher
      amplitudes)} near frame 40,000 {\color{red} at $\sim 2$ Hz,
      damping out to lower frequencies over time,} indicates periodic
    behavior corresponding with the systematic. The top and bottom
    0.1\% amplitudes have been masked out for visual clarity. This
    transform assumes the frames are evenly distributed in time, but
    the observations were taken in 64-frame chunks with relatively
    large separations.}
  \label{fig:wavelet}
\end{figure*}

The periodicity is localized in time, so we apply a continuous Morlet
wavelet transform, using the {\texttt{pywavelets}}
\citep{LeeEtal2019jossPyWavelets} Python package (see Figure
\ref{fig:wavelet}). Wavelet transforms assume a uniform sampling, but
our observations are sets of 64 short-cadence frames separated by
relatively long gaps, to work around data storage limits. This results
in spurious periodicity in the wavelet transforms. Despite this
limitation, a wavelet transform reveals periodic activity in the
elliptical area of the PRF at the time that the systematic occurred,
{\color{red} near frames 40,000 -- 50,000. In particular, there is a
  cluster of stronger amplitudes at $\sim 2$ Hz, damping out to lower
  frequencies over the course of several thousand frames.}

\begin{figure}
  \includegraphics[width=3.25in]{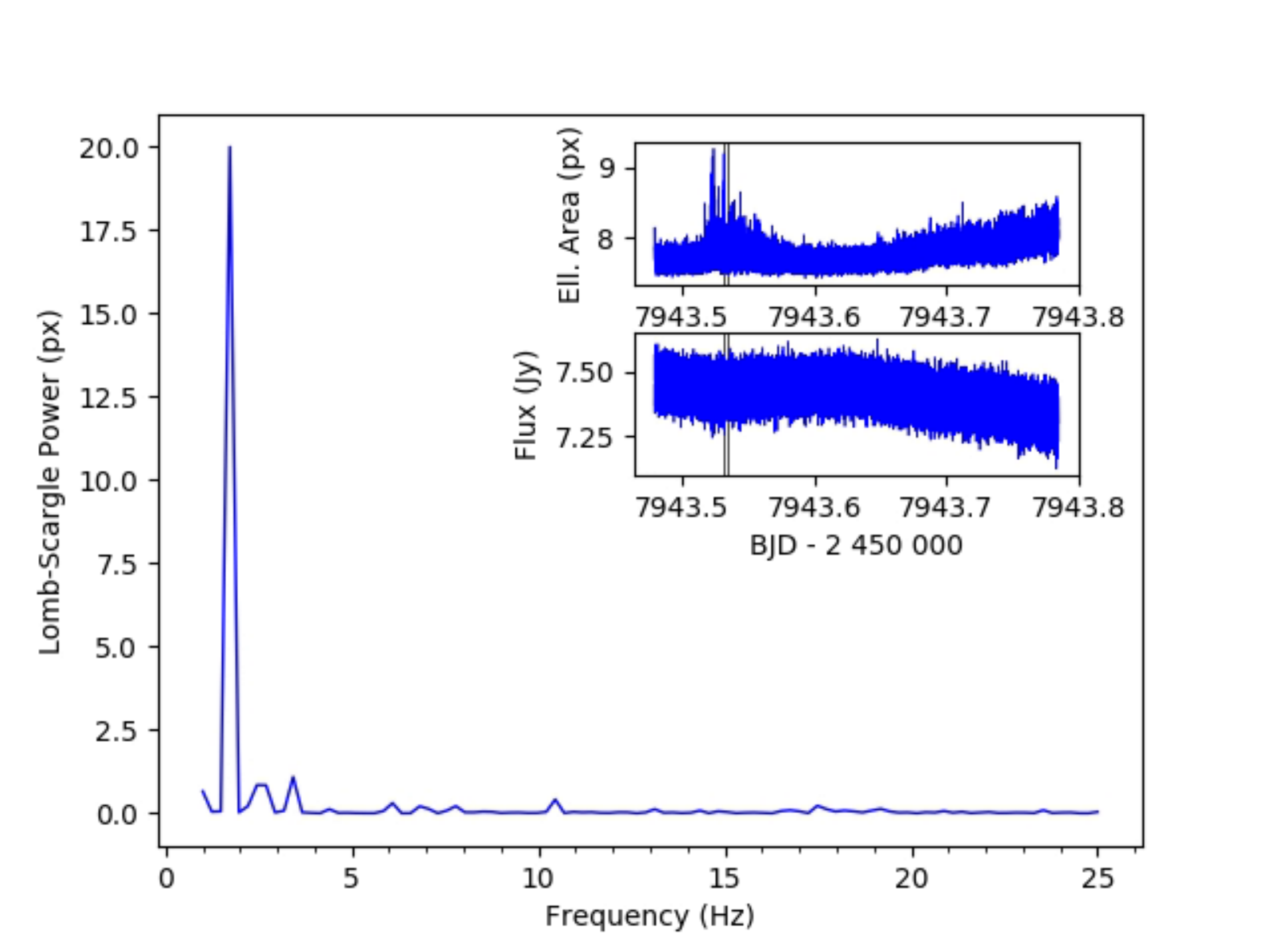}
  \caption{Windowed Lomb-Scargle periodogram of the elliptical area of
    the PRF during the July 2017 {\color{red} 3.6 \micron} observation. The insets show the
    elliptical area of the PRF and flux vs.\ time, with vertical black
    lines bracketing the five minute window used in the
    periodogram. The peak in power at $\sim$2 Hz, near the 0.45 second
    periodicity, appears during the systematic. Videos of this plot,
    using a sliding window, are available in the compendium.}
  \label{fig:ls}
\end{figure}

Lomb-Scargle periodograms are well-suited to finding periodicity in
non-uniformly sampled data, but unlike wavelet transforms, they
provide no temporal resolution of localized activity. {\color{red} To
  gain some insight into local periodicity, we use a windowed
  Lomb-Scargle periodogram} (see Figure \ref{fig:ls}). The periodogram
shows a strong peak at $\sim$ 2 Hz (as well as several weaker
resonances), which matches the periodic behavior seen in Figure
\ref{fig:sinefit}.

Until now, we calculated elliptical area from the $x$ and $y$ widths
of the PRF. However, this measure of area is only accurate if
vibrations are oriented along those axes. To more accurately measure
the shape of the PRF, we rotate a Gaussian clockwise from the $x$
axis. This detaches the $x$ and $y$ widths from their respective axes,
instead measuring the semimajor and semiminor axes of the ellipse. A
rotated 2D Gaussian is described by

\begin{equation}
  \label{eqn:rotgauss}
  \begin{split}    
    G(x,y,\sigma_{x'},\sigma_{y'},\theta,H) = H \textrm{exp}(&-g_1(x-x_0)^2\\
    &+2g_2(x-x_0)(y-y_0)\\
    &+g_3(y-y_0)^2)\\
    &+C
  \end{split}
\end{equation}

\noindent
where

\begin{align*}
    g_1&= \frac{\cos^2  \theta}{2\sigma_{x'}^2}+\frac{\sin^2  \theta}{2\sigma_{y'}^2}, \\
    g_2&=-\frac{\sin^2 2\theta}{4\sigma_{x'}^2}+\frac{\sin^2 2\theta}{4\sigma_{y'}^2}, \\
    g_3&= \frac{\sin^2  \theta}{2\sigma_{x'}^2}+\frac{\cos^2  \theta}{2\sigma_{y'}^2}, 
\end{align*}

\begin{figure}
  \includegraphics[width=3.25in, trim=0 50 0 50, clip]{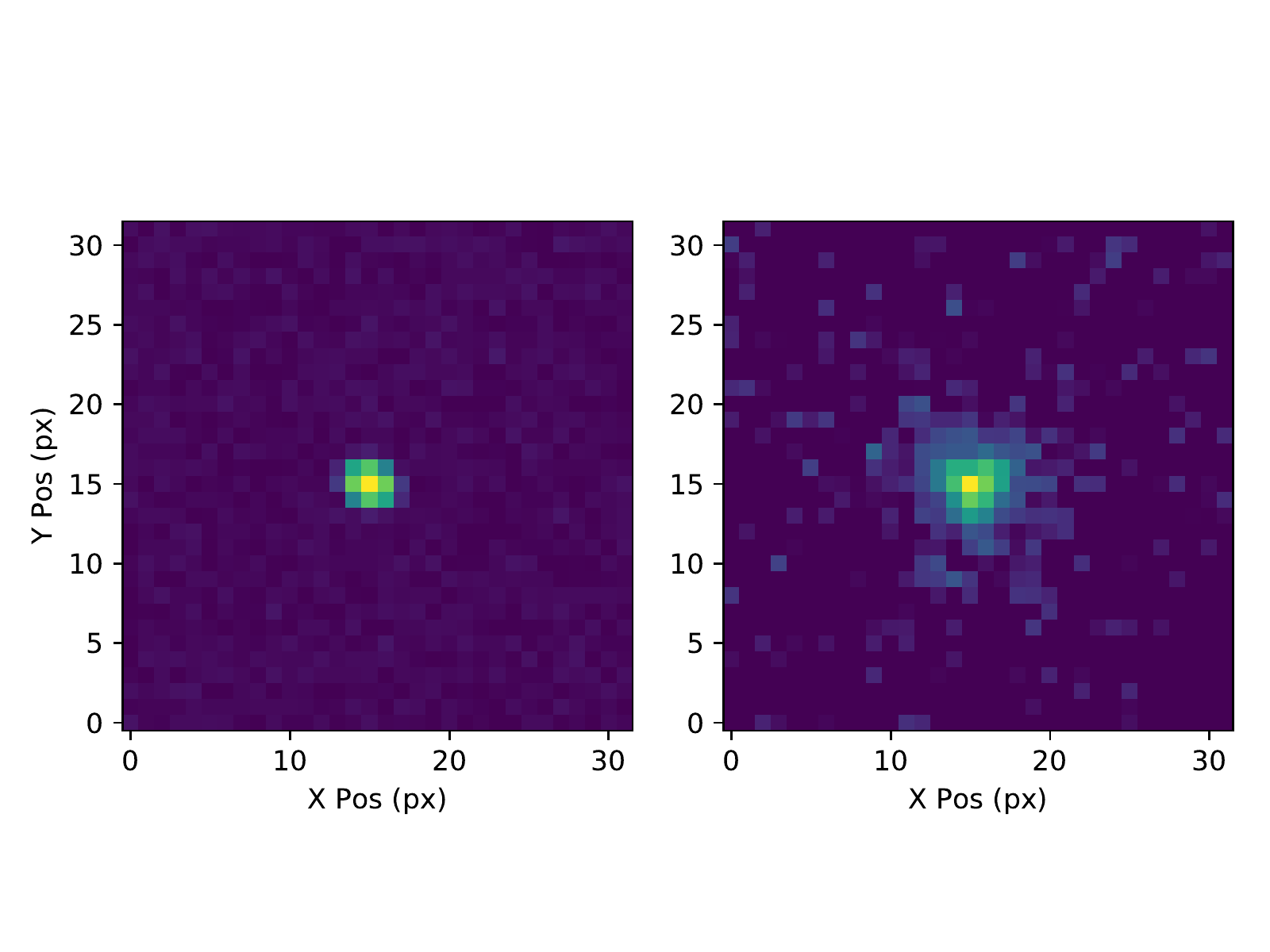}
  \caption{Log-scaled test images for the rotated, elliptical Gaussian
    centering. {\bf Left:} A synthetic image computed from Equation
    \ref{eqn:rotgauss} with Poisson noise. {\bf Right:} A real
    \textit{Spitzer} image of Proxima from AOR 63273472.}
  \label{fig:rottests}
\end{figure}

\noindent
$H$ is the height of the Gaussian, $\theta$ is the angle of rotation
clockwise from the $x$ axis, $x_0$ is the $x$ position of the peak,
$y_0$ is the $y$ position of the peak, $\sigma_{x'}$ is the width along
$\theta$, $\sigma_{y'}$ is the width along $\theta + 90^\circ$, and $C$
is a constant background level. We fit this Gaussian function to each
image using least-squares, weighted by the inverse of the
\textit{Spitzer}-supplied uncertainties, to determine the orientation
and shape of the PRF. We tested this algorithm on both a synthetic
rotated, elliptical Gaussian and an image from our observations (see
Figure \ref{fig:rottests}). The results are listed in Table
\ref{tbl:rottests}. The difference in retrieved star position is small
but differences in the measured PRF widths are more significant.

\begin{table*}[!ht]
\movetableright=-0.3in
\caption{\label{tbl:rottests}Rotated Gaussian Tests}
\begin{tabular}{lccccccccc}
  \hline
  \hline
Method & $\sigma_x$ & $\sigma_y$ & $\sigma_{x'}$ & $\sigma_{y'}$ & $x_0$ & $y_0$ & $H$ & $\theta$ & Background\\ 
\hline
\multicolumn{8}{l}{Synthetic Image}\\
\hline
Truth             & ---   & ---   & 0.600 & 0.500 & 15.000 & 15.000 & 87000 & $\pi/6$ (0.524) & 100.0\\
Std. Gaussian     & 0.570 & 0.521 & ---   & ---   & 15.004 & 15.000 & 87135 & ---             & 101.2\\
Rot. Gaussian     & ---   & ---   & 0.599 & 0.499 & 15.004 & 15.000 & 87187 & 0.526           & 100.2\\
\hline
\multicolumn{8}{l}{\textit{Spitzer} Image from AOR 63273472}\\
\hline
Std. Gaussian & 0.568 & 0.527 & ---   & ---   & 15.107 & 14.892 & 82175 & ---   & 32.7\\
Rot. Gaussian & ---   & ---   & 0.585 & 0.502 & 15.103 & 14.883 & 84120 & 0.508 & 32.9\\
\hline

\hline
\end{tabular}
\end{table*}

\begin{figure*}
  \includegraphics[width=2.33in]{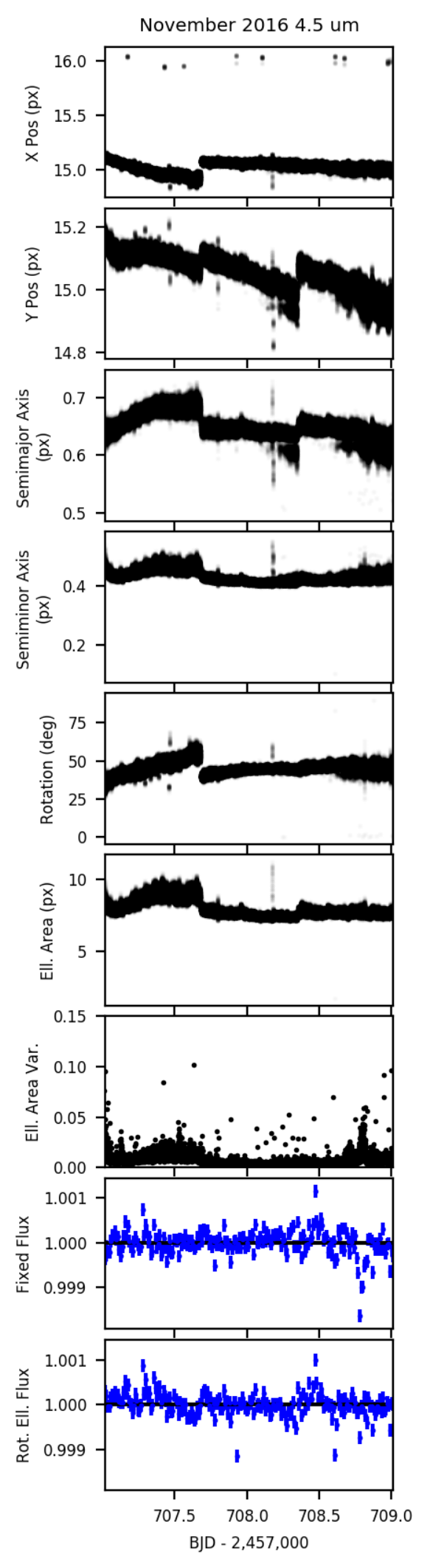}
  \includegraphics[width=2.33in]{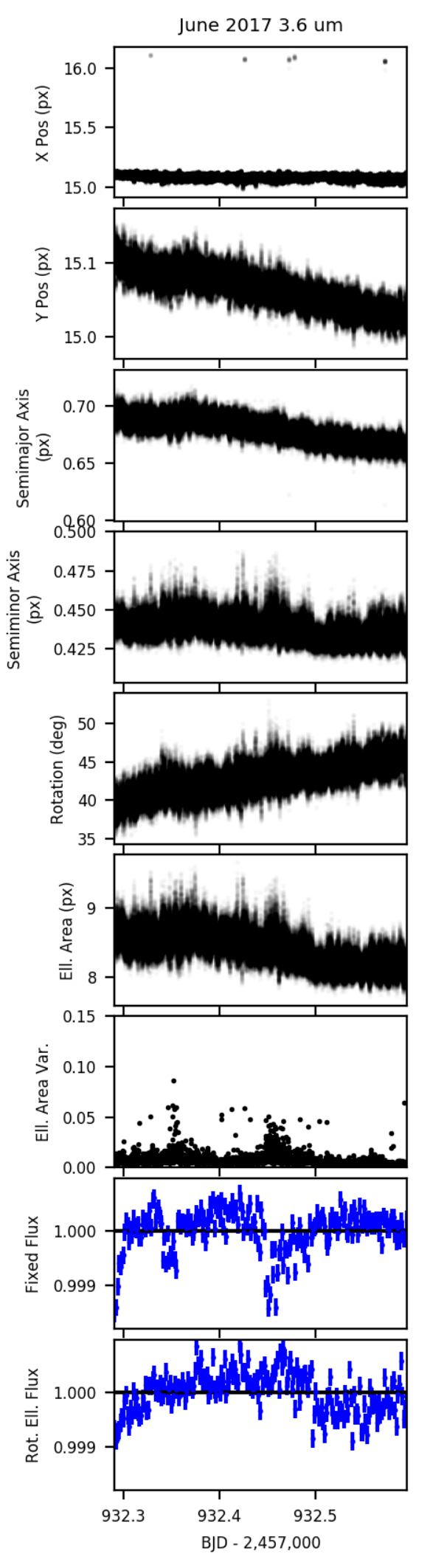}
  \includegraphics[width=2.33in]{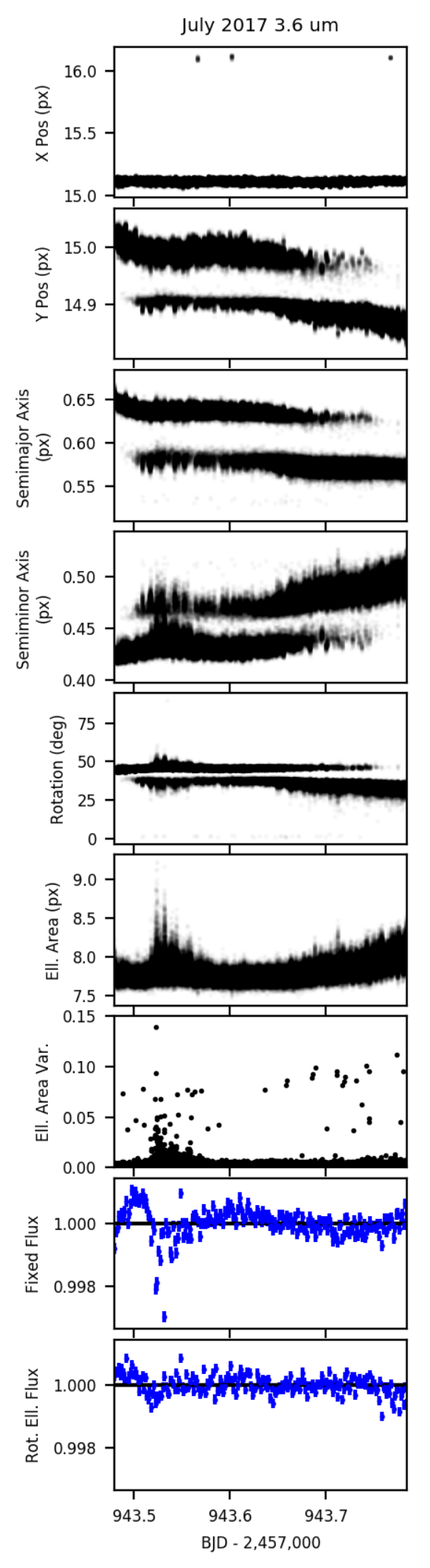}
  \caption{Results of rotated elliptical centering and
    photometry. Columns from left to right are the November 2016
    {\color{red} 4.5 \microns}, June 2017 {\color{red} 3.6 \microns},
    and July 2017 {\color{red} 3.6 \microns} observations. From top to
    bottom, rows are $x$ position, $y$ position, semimajor axis,
    semiminor axis, rotation, elliptical area, elliptical area
    variance, {\color{red} fixed-aperture light curve, and elliptical-aperture
    light curve.}}
  \label{fig:rfgc}
\end{figure*}

We applied this rotated-Gaussian centering method to the observations
affected by the systematic. The results are displayed in the first
seven rows of Figure \ref{fig:rfgc}. They match the non-rotated
Gaussian fits in elliptical area and elliptical area variance. These
systematic identification methods perform nearly equivalently when
using the non-rotated Gaussian. However, the rotated Gaussian has
implications for elliptical photometry, which is discussed in Section
\ref{sec:treat}.

\begin{figure}
  \includegraphics[width=3.5in, trim=50 0 50 0, clip]{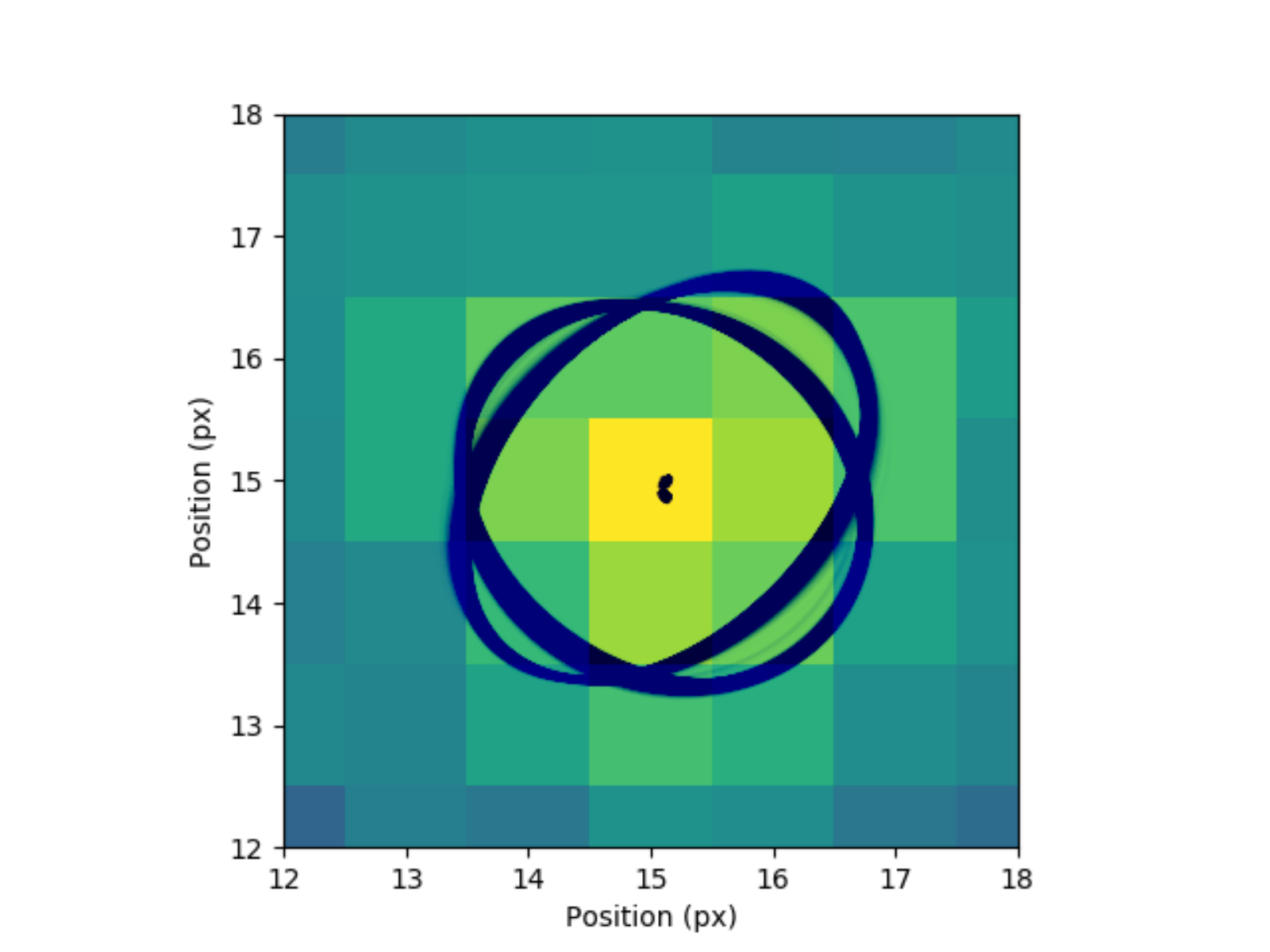}
  \caption{All $3\sigma$ rotated elliptical apertures for the July
    2017 {\color{red} 3.6 \micron} observation and their centers,
    overlaid on a log-scaled \textit{Spitzer} image from the same
    observation. The ellipses toggle between two rotational modes
    corresponding with the bimodal distribution in centering
    position.}
  \label{fig:ellvis}
\end{figure}

There are bimodalities in the fitted $y$ position, the axes lengths,
and rotation of the ellipse when the center of the PRF passes below
the center of a pixel. This behavior may be due to the asymmetry of
the IRAC PRF, which has a roughly-triangular shape (e.g., the second
panel of Figure \ref{fig:rottests}). The ellipse is swapping between
the asymmetric edges of the triangle (Figure \ref{fig:ellvis}). We see
this behavior in the Proxima images and the synthetic images created
with IRACSIM for the \textit{Spitzer} data challenge
\citep{IngallsEtal2016ajSpitzerSystematics}, but not with simple
synthetic Gaussians (Figure \ref{fig:rottests}), suggesting it is a
real effect of the complex PRF.

\section{SYSTEMATIC REMOVAL THROUGH ELLIPTICAL PHOTOMETRY}
\label{sec:treat}

Past works have removed this vibrational systematic prior to modeling
with variable, circular apertures
\citep[e.g.,][]{LewisEtal2013apjHATP2b, DemingEtal2015apjHATP20pld,
  GarhartEtal2018aapQatar1b, JenkinsEtal2019mnrasProxNoTran}. These
apertures adjust to avoid spilling light. However, due to their
circular shape, they must either spill flux from the aperture or
overcompensate in size for the elliptically-smeared PRF to capture all
the important pixels; thus, they include unnecessary background noise.

Instead, we use elliptical photometry, where we use an elliptical
aperture described by the fitted parameters from the non-rotated
Gaussian or rotated-Gaussian centering methods described in Section
\ref{sec:diag}. With rotated Gaussian centering, we apply the rotation
to the elliptical aperture. Similar to using variable-aperture
photometry, elliptical apertures attempt to remove the effects of PRF
activity prior to modeling, but only including the most important
pixels, resulting in less noise. Several elliptical photometry
packages exist \citep[e.g.,][]{LaherEtal2012paspAPT,
  Barbary2016jossSEP, MerlinEtal2019aaAPHOT}, although application has
been limited to correcting atmospheric effects in ground-based
observations \citep{BowmanHoldsworth2019aaEllPhot}, measuring the
radial surface brightness profiles of physically elliptical galaxies
\citep[e.g.,][]{DavisEtal1985ajGalSurfPhot, Djorkovski1985PhD,
  Cornell1989PhD, Ryder1992aujphGalaxies,
  McNamaraOConnell1992apjGalColorGrad, HayesEtal2005aaStarburst}, and
measuring photometry of comets that move significantly during each
exposure \citep{Miles2009sassEllComet}. To our knowledge, none have
used elliptical apertures to correct for vibrational effects.

\begin{figure*}
  \includegraphics[width=\textwidth]{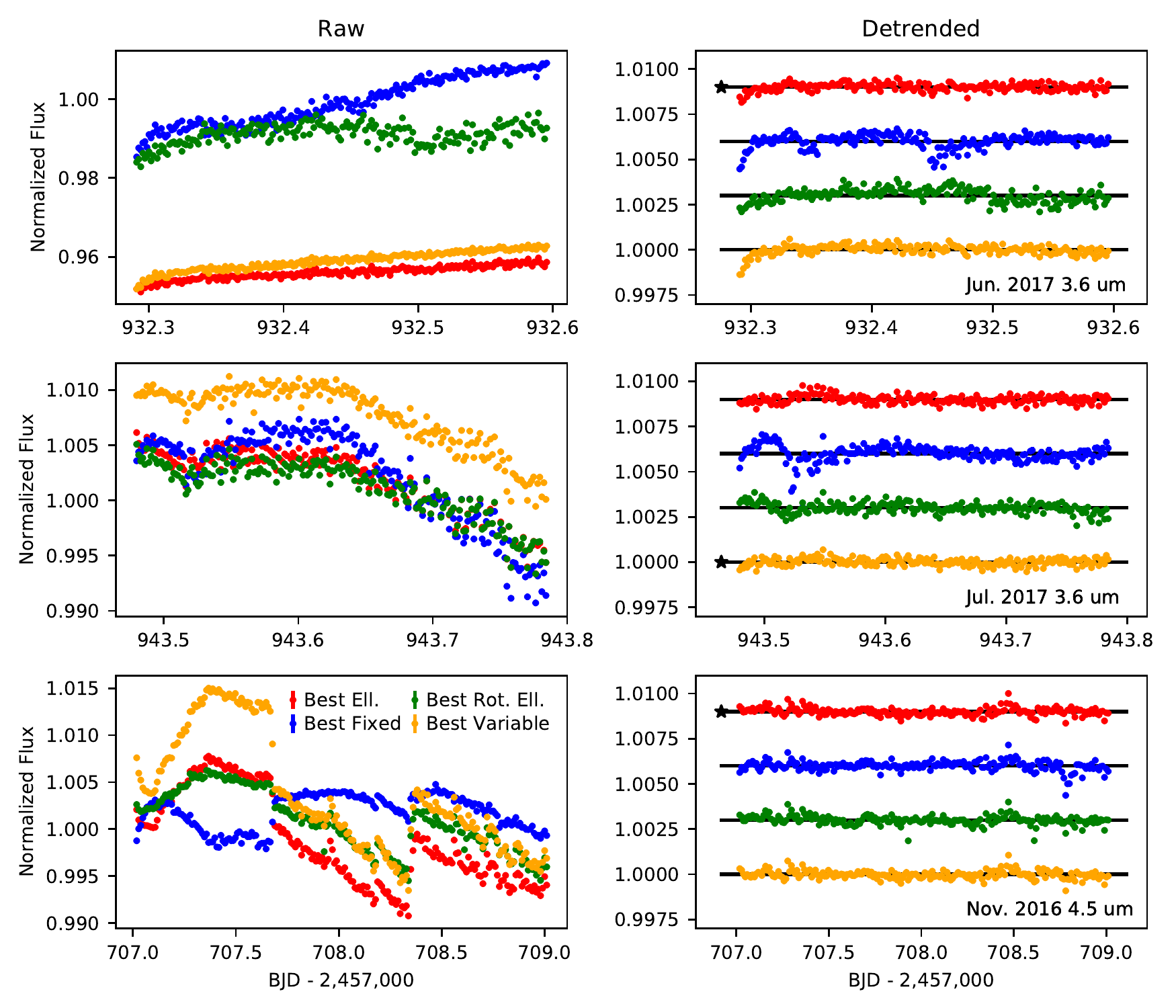}
  \caption{\label{fig:optphot} {\color{red} The best (minimum
      $\chi^2_{\textrm bin}$) raw, normalized, binned photometry
      (left) and BLISS-detrended light curves (right) for each
      photometry method. We have also divided out the time-dependent
      ramp models. In most cases, residual ramps are likely due to
      ramp models attempting to remove vibrational effects or the
      light-curve features introduced by the rotational bimodality in
      rotated elliptical aperture photometry, rather than the ramp
      effect. With elliptical photometry, where these non-ramp effects
      are reduced or non-existent, the minor residual ramp effect is
      likely due to differences in shape between our ramp models
      (flat, linear, or quadratic) and the ramp effect in that
      observation. The best (minimum $\chi^2_{\textrm bin}$) overall
      photometry (considering fixed-radius, variable-radius,
      elliptical, and rotated elliptical apertures) is marked with a
      black star.} See Table \ref{tbl:optphot} for the specific
    {\color{red} aperture sizes} used (in bold).}
\end{figure*}

Qualitatively, we find that elliptical photometry almost entirely
removes the vibrational systematic from the light curve, with the
non-rotated ellipses outperforming the rotated ones (see Figures
\ref{fig:identmeth} and \ref{fig:rfgc}, last rows). To assess
performance quantitatively, we fit BLISS and PLD models to the three
observations which include the vibrational systematic.  PLD performs
poorly when applied to observations longer than typical eclipses and
transits \citep{DemingEtal2015apjHATP20pld}, so, for the 48-hour
observation, we only consider the final 16 hours (after the final data
downlink). Many PLD implementations also bin the data
\citep[e.g.,][]{DemingEtal2015apjHATP20pld,WongEtal2015apjWASP14,
  BuhlerEtal2016apjHAT13b}, which can reduce short-period correlated
noise. We choose not to bin to isolate each model's ability to address
correlated noise. Table \ref{tbl:optphot} lists the results:
$\chi^2_{\textrm bin}$-minimized photometry aperture sizes for each
combination of centering and photometry methods, as well as the
$\chi^2_{\textrm bin}$ (lowest in bold) and SDNR for each combination,
both for BLISS and PLD fits. Figure \ref{fig:optphot} shows the
$\chi^2_{\textrm bin}$-minimized BLISS light curves compared with the
fixed-radius aperture light curves in Figure \ref{fig:subplots}.  We
reduced the strength of the systematic from 0.16\% to 0.06\% in the
November 2016 {\color{red} 4.5 \micron} observation, from 0.14\% to
0.06\% in the June 2017 {\color{red} 3.6 \micron} observation, and
from 0.30\% to 0.03\% in July 2017 {\color{red} 3.6 \micron}
observation, measured by the minimum of the binned photometry
presented in Figure \ref{fig:optphot}.

\begin{deluxetable*}{llrrrrrr}
  \tablecaption{\label{tbl:optphot}Optimal Photometry Methods}
  \tablehead{& & \multicolumn{3}{c}{BLISS} & \multicolumn{3}{c}{PLD} \\
    \colhead{Photometry} & \colhead{Centering} & \colhead{Ap. Size\tablenotemark{a}} & \colhead{$\chi^2_{\textrm bin}$} & \colhead{SDNR} & \colhead{Ap. Size\tablenotemark{a}} & \colhead{$\chi^2_{\textrm bin}$} & \colhead{SDNR}\\
    & & \colhead{(pixels)} & & \colhead{(ppm)} & \colhead{(pixels)} & & \colhead{(ppm)}
  }
  \tablecolumns{8}
  \startdata
  \hline
  \hline
\multicolumn{5}{l}{November 2016 {\color{red} 4.5 \micron} (last 16 hours of observation)}\\
\hline
\hline
Fixed      & Gaus.     & 3.00     & 21.8  & 7630 & 3.50     & 62.1  & 8173\\         
           & L. Asym.  & 3.00     & 22.1  & 7641 & 3.25     & 61.2  & 7909\\         
           & C. of L.  & 4.00     &293.2  & 8758 & 3.50     & 61.4  & 8169\\         
Variable   & Gaus.     & 0.50+1.0 &  7.7  & 7583 & 0.75+2.0 & \textbf{34.7}  & 8506\\
           & L. Asym.  & 0.50+1.0 &  9.0  & 7475 & 0.75+2.0 & 37.0  & 8509\\         
           & C. of L.  & 1.50+0.5 &200.3  & 8324 & 0.75+1.5 & 35.6  & 7932\\         
Elliptical & Gaus.     & 4.00+0.5 &  \textbf{5.1}  & 7438 & 3.00+2.0 & 42.5  & 8216\\         
           & Rot. Gaus.& 3.00+1.0 &  5.1  & 7727 & 5.00+1.5 & 60.3  & 8858\\         
\hline
\hline
\multicolumn{5}{l}{June 2017 {\color{red} 3.6 \micron}}\\
\hline
\hline
Fixed      & Gaus.     & 3.25     &  58.8 & 5511 & 3.00     &  74.2 & 5375\\         
           & L. Asym.  & 3.75     & 124.3 & 5778 & 2.75     &  76.4 & 5286\\         
           & C. of L.  & 4.50     &1440.0 & 6642 & 3.25     &  73.2 & 5490\\         
Variable   & Gaus.     & 0.75+0.5 &  12.5 & 5632 & 0.75+1.0 &  36.6 & 5417\\         
           & L. Asym.  & 1.00+0.0 &  21.0 & 5657 & 0.75+0.5 &  33.6 & 5503\\         
           & C. of L.  & 0.50+0.5 & 150.0 & 6627 & 0.75+0.5 &  \textbf{28.8} & 5375\\
Elliptical & Gaus.     & 4.00+0.0 &   \textbf{3.1} & 5295 & 5.00-0.5 &  29.3 & 5232\\         
           & Rot. Gaus.& 3.00+0.0 &   7.7 & 5808 & 6.00-0.5 &  31.3 & 5332\\         
\hline
\hline
\multicolumn{5}{l}{July 2017 {\color{red} 3.6 \micron}}\\ 
\hline
\hline
Fixed      & Gaus.     & 4.50     &  87.8 & 5926 & 4.00     & 159.0 & 5763\\         
           & L. Asym.  & 4.50     &  30.0 & 5889 & 4.00     & 159.3 & 5754\\         
           & C. of L.  & 4.50     &1175.9 & 6295 & 4.50     & 161.9 & 5982\\         
Variable   & Gaus.     & 1.50-0.5 &   2.6 & 5585 & 1.50-0.5 &  55.3 & 5582\\         
           & L. Asym.  & 1.00+0.5 &   \textbf{2.5} & 5437 & 1.00+0.5 &  56.9 & 5443\\         
           & C. of L.  & 0.50+0.0 &  45.3 & 8682 & 0.75+0.5 &  \textbf{36.9} & 5577\\
Elliptical & Gaus.     & 7.00-1.0 &   4.9 & 5229 & 7.00-1.0 &  71.1 & 5223\\         
           & Rot. Gaus.& 5.00+0.0 &  23.7 & 5225 & 7.00+0.5 &  93.2 & 5803\\         
\enddata
\tablenotetext{a}{Aperture sizes for variable and elliptical photometry are listed as $a+b$ (see Equations \ref{eqn:varrad} and \ref{eqn:ellrad}).}
\end{deluxetable*}

{
\color{red}
\section{COMPARISON TO GAUSSIAN-WIDTH DETRENDING}

Some analyses of \textit{Spitzer} IRAC data found that light-curve
correlated noise could be significantly reduced by combining a BLISS
map with a model dependent on the Gaussian widths of the PRF
\citep[e.g.,][]{LanotteEtal2014aapGJ436b, MendoncaEtal2018ajWASP43b,
  MansfieldEtal2020apjKELT9b}. Here we compare that approach with
elliptical-aperture photometry, and test the results of combining
the two methods.

For the Gaussian-width detrending, we use the following generic
quadratic model:

\begin{equation}
  \label{eqn:quadsig}
  W(\sigma_x, \sigma_y) = c_1\sigma_x^2 + c_2\sigma_y^2 + c_3\sigma_x\sigma_y + c_4\sigma_x + c_5\sigma_y + c_6,
\end{equation}

\noindent
where $c_i$ are free parameters in the light-curve model and
($\sigma_x$, $\sigma_y$) are the widths of the Gaussian fit to the PRF
of each data frame. We set $c_6 = 1$. This model is included in the
full light-curve model as a multiplicative factor on the right side of
Equation \ref{eq:bliss}.

We repeat the process used in Section \ref{sec:treat} of optimizing
for the minimum $\chi^2_{\textrm{bin}}$ in each photometry method,
although we are restricted to Gaussian centering, as the Gaussian
widths are an input to the light-curve model. A sample of the results
is shown in Figure \ref{fig:modelcomp}. Unsurprisingly, models that
include $W$ lead to better fits (lower $\chi^2$) due to increased
flexibility. In most cases, the noise reduction is clearly significant
when applied to fixed-radius aperture photometry. However, the
improvement is more marginal when the model is applied to non- and
rotated elliptical apertures. For example, for the July 2017 3.6
\microns\ observation, $\chi^2_{\textrm{bin}}$ for non-rotated
elliptical apertures improves from 4.9 to 4.8 and for rotated
elliptical apertures improves from 23.7 to 19.5. In some cases
including $W$ actually leads to more correlated noise because
model-fitting seeks to minimize $\chi^2$, not $\chi^2_{\textrm{bin}}$.
For all three observations, elliptical apertures led to a lower
$\chi^2_{\textrm{bin}}$ than fixed-radius circular apertures with a
Gaussian-width model.

We cannot use a statistic like the Bayesian Information Criterion
\citep{Raftery1995BIC} to decide if $W$ is worth including. The
$\chi^2_{\textrm{bin}}$-optimized photometry while including the $W$
model is different than the $\chi^2_{\textrm{bin}}$-optimized
photometry without $W$; we would be comparing model fits to different
data sets.

\begin{figure*}
  \includegraphics{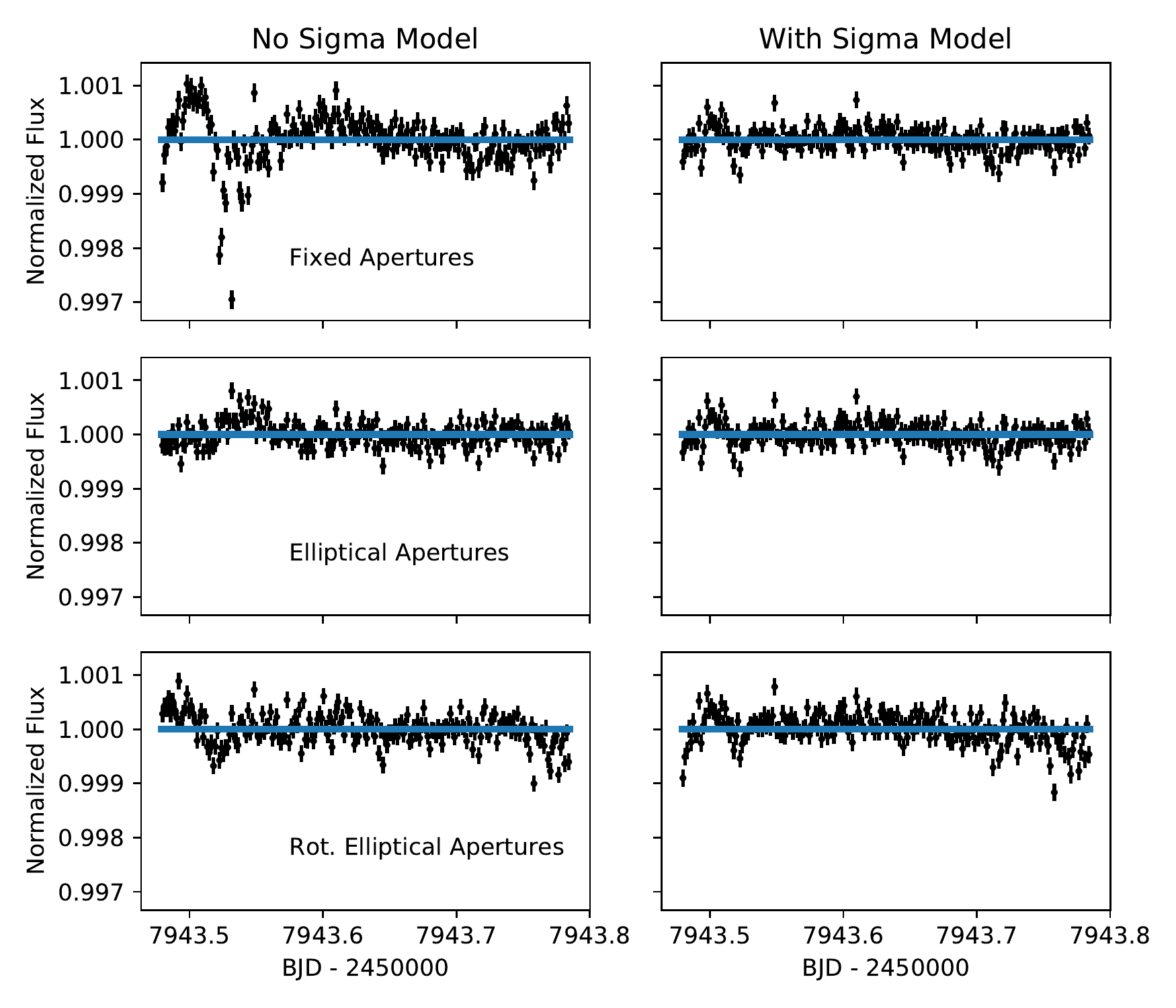}
  \caption{Comparison between the July 2017 3.6 \microns\ light curve
    detrended with (right) and without (left) the Gaussian-width model
    (Equation \ref{eqn:quadsig}). Rows are fixed-radius apertures,
    elliptical apertures, and rotated elliptical apertures.}
  \label{fig:modelcomp}
\end{figure*}
}

\section{DISCUSSION}
\label{sec:disc}

We draw several conclusions from the results in Table
\ref{tbl:optphot}.  First, we find elliptical photometry superior or
equivalent to variable, circular apertures when using BLISS maps. The
vibrational systematic is not correlated with position, especially if
the vibration occurs at a period shorter than the exposure time,
and thus cannot be corrected by a BLISS map. By removing the
vibrational systematic with elliptical photometry, the accuracy of the
BLISS map improves for the entire observation.

The PLD model is more flexible in its noise removal. It assumes that
flux variations are tied to fluctuations in the pixel brightnesses. As
the target moves on the detector, pixels brighten and dim. Likewise,
if the PRF is smeared, pixels near the center of the target dim and
pixels along the vibration axis brighten. Thus, the PLD model is able
to correct for the vibrational systematic without explicit knowledge
of the vibration, minimizing the advantage gained by using elliptical
photometry. This is convenient, but we achieve much lower correlated
noise in the BLISS models where the systematics are corrected with a
physical description of their effects (see $\chi^2_{\textrm{bin}}$
values in Table \ref{tbl:optphot}). We do not use binning in our
application of PLD, which would reduce correlated noise, but again
without explicit knowledge of the vibration. {\color{red} For example,
  when testing bin sizes of 1, 2, 4, 8, 18, 32, 64, 128, and 256 on
  the optimal PLD light curves from Table \ref{tbl:optphot}, we
  improve the $\chi^2_{\textrm{bin}}$ from 34.7 to 34.3 for the
  November 2016 4.5 \micron\ observation, from 28.8 to 12.0 for the
  June 2017 3.6 \microns\ observation, and from 36.9 to 5.2 for the
  July 2017 3.6 microns\ observation. This suggests that temporal
  binning provides a significant portion of PLD's correlated-noise
  correction capabilities.}

The rotated ellipse is never preferred over the non-rotated case. As
mentioned above, the \textit{Spitzer} PRF is highly asymmetric, and
slightly triangular in shape (see Figure \ref{fig:rottests}, right
panel), which creates a challenge when fitting a rotated ellipse. The
vibration-induced elliptical shape is less prominent than the
already-present asymmetry in the PRF, as evidenced by the bimodal
distribution in rotation (Figure \ref{fig:rfgc}). We suspect that the
rotated elliptical Gaussian is fitting to the sides of the triangular
PRF, which creates additional noise in the resulting light
curve. {\color{red} In particular, the bimodality creates problems for
  BLISS maps in two ways:

  \begin{enumerate}
  \item BLISS maps rely on multiple frames at the same location to
    accurately calculate subpixel sensitivity. The bimodality
    reduces the number of frames in each sensitivity map grid cell,
    leading to greater uncertainty.

  \item The bimodality increases the size of the grid cells, which
    we calculate as the RMS of the point-to-point target position
    variation in $x$ and $y$. This reduces the flexibility of the
    model, resulting in worse fits.
  \end{enumerate}

  }
\noindent
Rotated elliptical photometry may be useful for other telescopes that
have more circular PRFs.

Since the \textit{Spitzer} PRF is a complex shape, ideally we would
determine flux by directly fitting the PRF, but that has proven
challenging. The \textit{Spitzer} PRF is underresolved, especially at
shorter wavelengths, and the true PSF is not known at a high
resolution, only as a map of a point source at a 5$\times$5 grid of
positions within a central pixel. Hence, we recommend overresolved
PRFs for high-precision point-source instruments like exoplanet
telescopes, or a high-resolution lab-measured PRF tested in comparison
to real data with a routine to accurately bin to the native pixel
level. One could also fit a shape more representative of the PRF, like
a tri-lobed Gaussian with a radial scale, rotation, stretching factor,
and stretching axis. However, that is beyond the scope of this work.

In general, we find that PLD is agnostic to the centering method used.
In two observations, we prefer center-of-light centering, and in the
third there is no strong preference for any of the methods. This would
suggest that, when using PLD models, it is acceptable to only apply
center-of-light centering, although we recommend always applying all
methods available.

BLISS maps, on the other hand, are extremely sensitive to the
centering method because 1) target position is an input to the model,
and 2) we use BLISS map $x$ and $y$ grid sizes equal to the RMS of the
point-to-point $x$ and $y$ target position motion, respectively. Thus,
higher precision centering methods result in maps with finer
structure. Compared to Gaussian and least-asymmetry centering,
center-of-light centering results in high RMS of point-to-point $x$
and $y$ target position motion and, thus, poor maps, at least for 3.6
and 4.5 \microns\ observations (Table \ref{tbl:optphot}). Therefore,
center-of-light can be ignored with BLISS maps, although applying all
analysis methods will ensure the best is chosen.

{\color{red} While including a Gaussian-width model can significantly
  reduce noise in fixed-aperture photometry, the addition of this
  model only slightly improves elliptical-aperture photometry. We
  recommend elliptical apertures over a Gaussian-width model for two
  reasons: 1) to reduce model complexity, and 2) to correct for
  correlated noise in image processing, rather than creating it with
  circular apertures and then modeling it out.}

Finally, in nearly all cases, non-rotated elliptical photometry
results in the lowest SDNR. With BLISS, elliptical photometry improves
SDNR by up to 11.2\% over fixed circular apertures and up to 6.0\%
over variable, circular apertures. With PLD, we see up to 9.4\%
improvement over fixed apertures and up to 6.3\% improvement over
variable apertures. These statistics are for the entire modeled light
curve; the improvement is even more pronounced if we only consider
data when the systematic is present. {\color{red} This suggests SDNR
  can be useful when optimizing photometric extraction, with the
  caveat that the SDNR values listed in Table \ref{tbl:optphot} are
  for the $\chi^2_{\textrm{bin}}$-optimized light curves, not
  SDNR-optimized light curves. In our case, if we optimized solely for
  SDNR, we would be left with fixed-radius apertures for the November
  2016 4.5 \microns\ observation, which is clearly unsatisfactory (see
  Figure \ref{fig:subplots}, top panel).}

The optimal light curves presented here are available, in machine- and
human-readable formats, in a compendium archive available at
\dataset[https://doi.org/10.5281/zenodo.3759914]{https://doi.org/10.5281/zenodo.3759914}. The
compendium also includes best-fit models and correlated noise
diagnostics.


\section{RESULTS}
\label{sec:results}

We have identified a vibrational systematic in \textit{Spitzer}
photometry that mimics planetary or cometary transits. With our short
exposure times, we were able to resolve this vibration in the size and
shape of the PRF, both on sub-second timescales and with
periodograms. We caution against false positive detections of planets,
and recommend applying the techniques described here to identify and
correct the systematic.

``Noise pixels'' can occasionally identify this systematic, but
they can be misleading, as noise pixel activity does not always
correspond with the systematic, and can frequently be hidden
in the baseline activity. Several other metrics are better suited
to identifying this vibration:

\begin{enumerate}
  \item $x$ and $y$ widths from Gaussian centering, both rotated and
    non-rotated.
  \item Elliptical area of Gaussian centering, both rotated and non-rotated.
  \item Variance of noise pixels.
  \item Variance of elliptical area.
  \item Wavelet amplitude over a variety of frequencies.
  \item Lomb-Scargle periodograms of elliptical area.
\end{enumerate}

\noindent
For our observations, variance of the PRF area most accurately
identifies the systematic. However, in most IRAC time-series
observations, identification of this systematic is more challenging.
The pointing wander induced by temperature fluctuation in the
telescope reduces the clarity of our diagnostics.

To correct this vibrational systematic, we developed an adaptive
elliptical-photometry technique. We fit an asymmetric Gaussian to the
PRF to determine target position and PRF shape, and use this
parameterization to create an elliptical aperture that adapts its
shape to the PRF as it changes with time. We applied elliptical
photometry to three observations known to include the vibrational
systematic, with both BLISS and PLD models to assess relative
performance. With BLISS models, elliptical photometry results in
reduced correlated noise in two of our three observations, and reduced
SDNR in all observations. PLD prefers variable, circular apertures
over elliptical apertures, but, without binning, is less capable of
removing correlated noise compared to BLISS. We also used a rotated
elliptical aperture, but found that the complex shape of the
\textit{Spitzer} PRF created bimodalities in the orientation of the
ellipse and noise in the resulting light curve. Other shapes, like a
tri-lobed Gaussian, are an area of potential future
study. {\color{red} Finally, we found that elliptical apertures
  outperformed traditional fixed-radius circular apertures with a
  Gaussian-width model.}

We cannot determine the source of the vibration, though we speculate
that it could be micrometeorite impacts or wear-and-tear on the
telescope, such as a defect in the gyroscopes. If the source is
micrometeorites, this systematic should be present in many past
observations, at roughly the same rate as in our observations (four
instances in 80 hours). Reanalyses with our techniques may be able to
rescue data sets deemed unsalvageable, or at least improve the
uncertainties on measured planetary transmission, emission, and phase
curve variation. If wear-and-tear is the source of the systematic,
then older observations may be unaffected, but more recent observations
would still be affected. \textit{Spitzer} produced high-profile
exoplanet science for 16 years
\citep[e.g.,][]{GillonEtal2017natTRAPPIST,
  KreidbergEtal2019natLHS3844b}, much of which is done at the limit of
detection. Elliptical photometry could make the difference between
speculation and discovery.

Elliptical photometry is not limited to
\textit{Spitzer}. TESS and \textit{Kepler} (and K2) are
purely photometric observatories that may suffer from the same
systematic. JWST also has photometric modes which will surely be used
to push transiting exoplanet photometry to the smallest and coldest
objects possible. Optimistically assuming that we reach the noise
floor, we will need large amounts of JWST time to study these planets
\citep[e.g.,][]{MorleyEtal2017apjJWSTearth}, and require the absolute
best data reduction and noise removal techniques.

\acknowledgments

We thank contributors to SciPy, Matplotlib, and the Python Programming
Language, the free and open-source community, the NASA Astrophysics
Data System, and the JPL Solar System Dynamics group for software and
services.  This work is based on observations made with the {\em
  Spitzer Space Telescope}, which was operated by the Jet Propulsion
Laboratory, California Institute of Technology under a contract with
NASA. The authors acknowledge support from the following:
CATA-Basal/Chile PB06 Conicyt and Fondecyt/Chile project \#1161218
(JSJ). Spanish MINECO programs AYA2016-79245-C03-03-P,
ESP2017-87676-C05-02-R (ER), ESP2016-80435-C2-2-R (EP) and through the
``Centre of Excellence Severo Ochoa'' award SEV-2017-0709 (PJA, CRL
and ER). STFC Consolidated Grant ST/P000592/1 (GAE). NASA Planetary
Atmospheres Program grant NNX12AI69G and NASA Astrophysics Data
Analysis Program grant NNX13AF38G (RC, JH, KM, MH). Spanish Ministry
of Science, Innovation and Universities and the Fondo Europeo de
Desarrollo Regional (FEDER) through grant ESP2016-80435-C2-1-R and
PGC2018-098153-B-C33 (IR).\\
\\
\facility{Spitzer (IRAC)}

\software{NumPy \citep{HarrisEtal2020natNumPy}, Matplotlib
  \citep{Hunter2007cseMatplotlib}, SciPy
  \citep{VirtanenEtal2020natmSciPy}, MC$^3$
  \citep{CubillosEtal2017ajRedNoise}, POET
  \citep[e.g.,][]{HardyEtal2017apjHATP13b}}

\appendix
\section{Optimizing Data Sets with $\chi^2_{\rm bin}$}
\label{app:bsig}

\begin{figure}[h]  
  \includegraphics[width=3.5in]{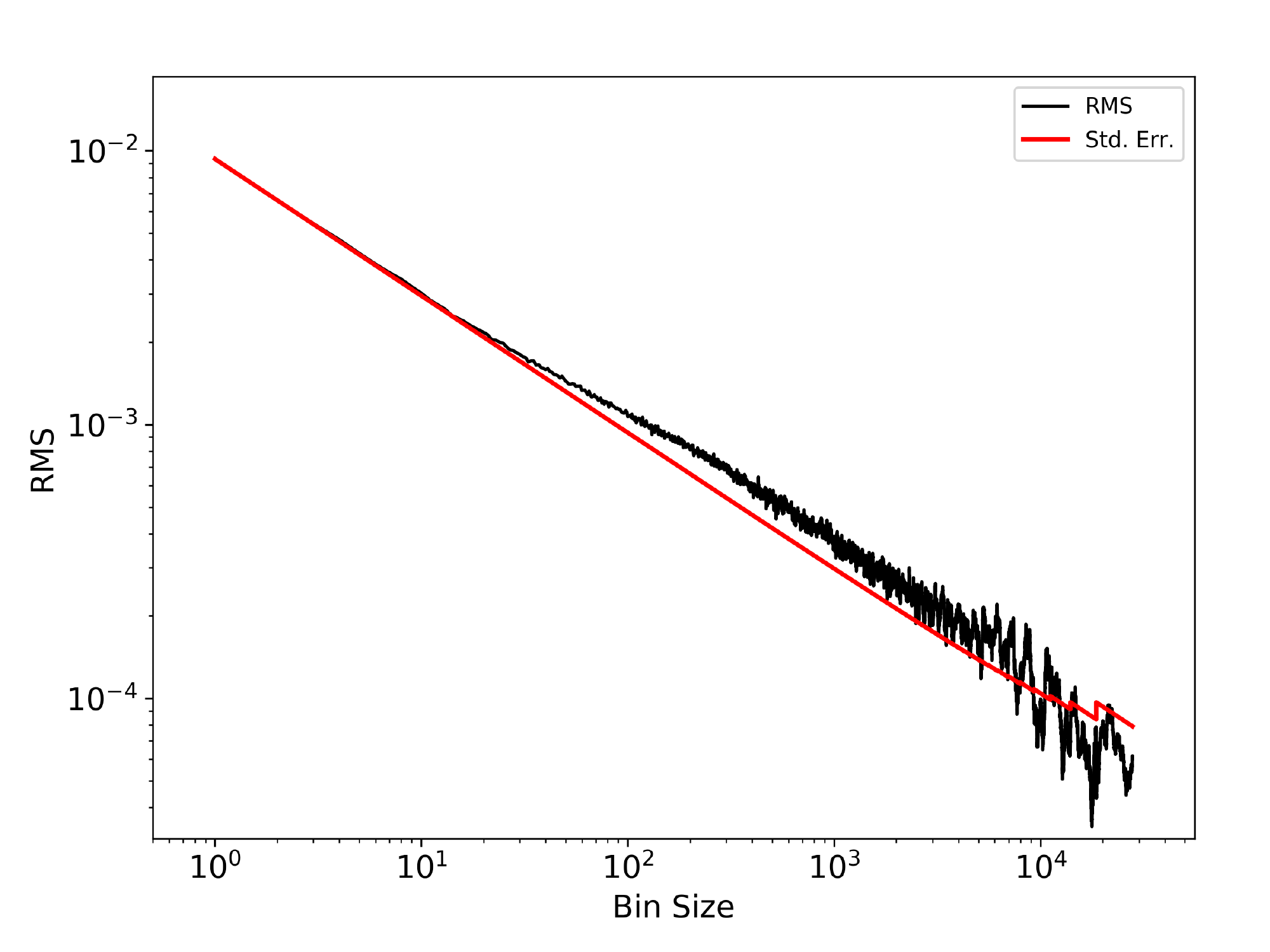}
  \caption{\label{fig:rmsexample}A plot of the
    root-mean-squared (RMS) of light-curve model residuals
    vs.\ residual bin size compared to theory, assuming only
    non-correlated noise. Correlated noise is present in the data at a
    given binning level if the black line (measured RMS) lies above
    the red line (theoretical RMS).}
\end{figure}

In this work, we choose the optimal centering methods, photometry
techniques, and photometry aperture sizes by minimizing $\chi^2_{\rm
  bin}$, a measurement of residual correlated noise
\citep{DemingEtal2015apjHATP20pld}. Here we describe that calculation
in detail. This calculation assesses correlated noise like a
root-mean-square vs.\ bin size plot (see Figure \ref{fig:rmsexample})
but in a more quantifiable way.

First, we define the standard deviation of normalized residuals (SDNR)
as

\begin{equation}
  {\rm SDNR} = \sqrt{\frac{1}{N-M}\sum_{i=0}^N \left(r_i - \bar r\right)^2},
\end{equation}

\noindent
where $r_i$ is the normalized model residual for frame $i$, $\bar r$
is the mean of the normalized residuals, $N$ is the number of frames,
and $M$ is the number of free parameters in the light-curve model.
Normalized model residuals are given by

\begin{equation}
  r = \frac{{\rm data} - {\rm model}}{\mathrm{planetless\ model}},
\end{equation}

\noindent
where the planetless model is the best-fitting model evaluated
without any planet terms (i.e., no eclipses, transits, or phase curve
variation). In this particular work, there are no planets, so this
is done by default.

If $r$ contains only white noise, then, when binned, SDNR should
decrease (improve) by a factor of $1/\sqrt{\rm bin\ size}$, where bin
size is the number of frames over which we average. On the other hand,
if there is correlated noise in $r$, binning will not improve the SDNR
as much.  Thus, we define an Expected SDNR (ESDNR) as

\begin{equation}
  {\rm ESDNR}_i = \frac{{\rm SDNR}_1}{\sqrt{i}},
\end{equation}

\noindent
where $i$ is the number of residual points per bin (bin size),
SDNR$_i$ is the SDNR with bin size $i$, and ESDNR$_i$ is the
ESDNR at bin size $i$.

We calculate a $\chi^2$ goodness-of-fit measurement for SDNR vs.\ bin
size compared to ESDNR, given by

\begin{equation}
  \label{eqn:bsig}
  \chi^2_{\rm bin} = \sqrt{\sum_{i=0}^n \left(\frac{{\rm SDNR}_{2^i} - {\rm ESDNR}_{2^i}}{\sigma_{{\rm SDNR}_{2^i}}}\right)^2},
\end{equation}

\noindent
where $n$ is the largest integer possible such that a bin size of
$2^n$ leaves more residual points than free parameters in the
light-curve model, but $2^{n+1}$ does not. $\sigma_{{\rm SDNR}_i}$ is
the uncertainty on SDNR$_i$, given by

\begin{equation}
  \sigma_{{\rm SDNR}_i} = \frac{{\rm SDNR}_1}{\sqrt{2N_{\rm bin}}},
\end{equation}

\noindent
where $N_{\rm bin}$ is the number of residual points left after
binning with bin size $i$. In Equation \ref{eqn:bsig} we bin by
factors of $2^i$, creating an evenly distributed number of bin sizes
in log space, to avoid biasing $\chi^2_{\rm bin}$ toward data sets
with less correlated noise at large bin sizes.

\bibliography{systematics.bib}

\begin{thebibliography}{}
\expandafter\ifx\csname natexlab\endcsname\relax\def\natexlab#1{#1}\fi
\providecommand{\url}[1]{\href{#1}{#1}}
\providecommand{\dodoi}[1]{doi:~\href{http://doi.org/#1}{\nolinkurl{#1}}}
\providecommand{\doeprint}[1]{\href{http://ascl.net/#1}{\nolinkurl{http://ascl.net/#1}}}
\providecommand{\doarXiv}[1]{\href{https://arxiv.org/abs/#1}{\nolinkurl{https://arxiv.org/abs/#1}}}

\bibitem[{{Anglada-Escud{\'e}} {et~al.}(2016){Anglada-Escud{\'e}}, {Amado},
  {Barnes}, {Berdi{\~n}as}, {Butler}, {Coleman}, {de La Cueva}, {Dreizler},
  {Endl}, {Giesers}, {Jeffers}, {Jenkins}, {Jones}, {Kiraga}, {K{\"u}rster},
  {L{\'o}pez-Gonz{\'a}lez}, {Marvin}, {Morales}, {Morin}, {Nelson}, {Ortiz},
  {Ofir}, {Paardekooper}, {Reiners}, {Rodr{\'{\i}}guez},
  {Rodr{\'{\i}}guez-L{\'o}pez}, {Sarmiento}, {Strachan}, {Tsapras}, {Tuomi}, \&
  {Zechmeister}}]{Anglada-EscudeEtal2016NatProxb}
{Anglada-Escud{\'e}}, G., {Amado}, P.~J., {Barnes}, J., {et~al.} 2016, \nat,
  536, 437, \dodoi{10.1038/nature19106}

\bibitem[{{Ballard} {et~al.}(2010){Ballard}, {Charbonneau}, {Deming},
  {Knutson}, {Christiansen}, {Holman}, {Fabrycky}, {Seager}, \&
  {A'Hearn}}]{BallardEtal2010paspGJ436c}
{Ballard}, S., {Charbonneau}, D., {Deming}, D., {et~al.} 2010, \pasp, 122,
  1341, \dodoi{10.1086/657159}

\bibitem[{{Barbary}(2016)}]{Barbary2016jossSEP}
{Barbary}, K. 2016, The Journal of Open Source Software, 1, 58,
  \dodoi{10.21105/joss.00058}

\bibitem[{{Batalha} {et~al.}(2018){Batalha}, {Lewis}, {Line}, {Valenti}, \&
  {Stevenson}}]{BatalhaEtal2018ApJjwsttemp}
{Batalha}, N.~E., {Lewis}, N.~K., {Line}, M.~R., {Valenti}, J., \& {Stevenson},
  K. 2018, \apjl, 856, L34, \dodoi{10.3847/2041-8213/aab896}

\bibitem[{{Beichman} \& {Greene}(2018)}]{Beichman&Greene2018haexJWST}
{Beichman}, C.~A., \& {Greene}, T.~P. 2018, {Observing Exoplanets with the
  James Webb Space Telescope} (Springer International Publishing), 85,
  \dodoi{10.1007/978-3-319-55333-7_85}

\bibitem[{{Blecic} {et~al.}(2013){Blecic}, {Harrington}, {Madhusudhan},
  {Stevenson}, {Hardy}, {Cubillos}, {Hardin}, {Campo}, {Bowman}, {Nymeyer},
  {Loredo}, {Anderson}, \& {Maxted}}]{BlecicEtal2013apjWASP14b}
{Blecic}, J., {Harrington}, J., {Madhusudhan}, N., {et~al.} 2013, \apj, 779, 5,
  \dodoi{10.1088/0004-637X/779/1/5}

\bibitem[{{Blecic} {et~al.}(2014){Blecic}, {Harrington}, {Madhusudhan},
  {Stevenson}, {Hardy}, {Cubillos}, {Hardin}, {Bowman}, {Nymeyer}, {Anderson},
  {Hellier}, {Smith}, \& {Collier Cameron}}]{BlecicEtal2014apjWASP43b}
---. 2014, \apj, 781, 116, \dodoi{10.1088/0004-637X/781/2/116}

\bibitem[{{Bowman} \& {Holdsworth}(2019)}]{BowmanHoldsworth2019aaEllPhot}
{Bowman}, D.~M., \& {Holdsworth}, D.~L. 2019, \aap, 629, A21,
  \dodoi{10.1051/0004-6361/201935640}

\bibitem[{{Buhler} {et~al.}(2016){Buhler}, {Knutson}, {Batygin}, {Fulton},
  {Fortney}, {Burrows}, \& {Wong}}]{BuhlerEtal2016apjHAT13b}
{Buhler}, P.~B., {Knutson}, H.~A., {Batygin}, K., {et~al.} 2016, \apj, 821, 26,
  \dodoi{10.3847/0004-637X/821/1/26}

\bibitem[{{Charbonneau} {et~al.}(2005){Charbonneau}, {Allen}, {Megeath},
  {Torres}, {Alonso}, {Brown}, {Gilliland}, {Latham}, {Mandushev}, {O'Donovan},
  \& {Sozzetti}}]{CharbonneauEtal2005apjTrES1}
{Charbonneau}, D., {Allen}, L.~E., {Megeath}, S.~T., {et~al.} 2005, \apj, 626,
  523, \dodoi{10.1086/429991}

\bibitem[{{Cornell}(1989)}]{Cornell1989PhD}
{Cornell}, M.~E. 1989, PhD thesis, Arizona Univ., Tucson.

\bibitem[{{Cubillos} {et~al.}(2017){Cubillos}, {Harrington}, {Loredo}, {Lust},
  {Blecic}, \& {Stemm}}]{CubillosEtal2017ajRedNoise}
{Cubillos}, P., {Harrington}, J., {Loredo}, T.~J., {et~al.} 2017, \aj, 153, 3,
  \dodoi{10.3847/1538-3881/153/1/3}

\bibitem[{{Cubillos} {et~al.}(2014){Cubillos}, {Harrington}, {Madhusudhan},
  {Foster}, {Lust}, {Hardy}, \& {Bowman}}]{CubillosEtal2014apjTrES1b}
{Cubillos}, P., {Harrington}, J., {Madhusudhan}, N., {et~al.} 2014, \apj, 797,
  42, \dodoi{10.1088/0004-637X/797/1/42}

\bibitem[{{Cubillos} {et~al.}(2013){Cubillos}, {Harrington}, {Madhusudhan},
  {Stevenson}, {Hardy}, {Blecic}, {Anderson}, {Hardin}, \&
  {Campo}}]{CubillosEtal2013ApjWASP8b}
---. 2013, \apj, 768, 42, \dodoi{10.1088/0004-637X/768/1/42}

\bibitem[{{Davis} {et~al.}(1985){Davis}, {Cawson}, {Davies}, \&
  {Illingworth}}]{DavisEtal1985ajGalSurfPhot}
{Davis}, L.~E., {Cawson}, M., {Davies}, R.~L., \& {Illingworth}, G. 1985, \aj,
  90, 169, \dodoi{10.1086/113723}

\bibitem[{{de Wit} {et~al.}(2012){de Wit}, {Gillon}, {Demory}, \&
  {Seager}}]{DeWitEtal2012aaHD189Map}
{de Wit}, J., {Gillon}, M., {Demory}, B.~O., \& {Seager}, S. 2012, \aap, 548,
  A128, \dodoi{10.1051/0004-6361/201219060}

\bibitem[{{Deming} {et~al.}(2015){Deming}, {Knutson}, {Kammer}, {Fulton},
  {Ingalls}, {Carey}, {Burrows}, {Fortney}, {Todorov}, {Agol}, {Cowan},
  {Desert}, {Fraine}, {Langton}, {Morley}, \&
  {Showman}}]{DemingEtal2015apjHATP20pld}
{Deming}, D., {Knutson}, H., {Kammer}, J., {et~al.} 2015, \apj, 805, 132,
  \dodoi{10.1088/0004-637X/805/2/132}

\bibitem[{{Deming} \& {Seager}(2017)}]{DemingSeager2017JGREreview}
{Deming}, L.~D., \& {Seager}, S. 2017, Journal of Geophysical Research
  (Planets), 122, 53, \dodoi{10.1002/2016JE005155}

\bibitem[{{Djorgovski}(1985)}]{Djorkovski1985PhD}
{Djorgovski}, S.~B. 1985, PhD thesis, California Univ., Berkeley.

\bibitem[{{Fazio} {et~al.}(2004){Fazio}, {Hora}, {Allen}, {Ashby}, {Barmby},
  {Deutsch}, {Huang}, {Kleiner}, {Marengo}, {Megeath}, {Melnick}, {Pahre},
  {Patten}, {Polizotti}, {Smith}, {Taylor}, {Wang}, {Willner}, {Hoffmann},
  {Pipher}, {Forrest}, {McMurty}, {McCreight}, {McKelvey}, {McMurray}, {Koch},
  {Moseley}, {Arendt}, {Mentzell}, {Marx}, {Losch}, {Mayman}, {Eichhorn},
  {Krebs}, {Jhabvala}, {Gezari}, {Fixsen}, {Flores}, {Shakoorzadeh}, {Jungo},
  {Hakun}, {Workman}, {Karpati}, {Kichak}, {Whitley}, {Mann}, {Tollestrup},
  {Eisenhardt}, {Stern}, {Gorjian}, {Bhattacharya}, {Carey}, {Nelson},
  {Glaccum}, {Lacy}, {Lowrance}, {Laine}, {Reach}, {Stauffer}, {Surace},
  {Wilson}, {Wright}, {Hoffman}, {Domingo}, \& {Cohen}}]{FazioEtal2004apjsIRAC}
{Fazio}, G.~G., {Hora}, J.~L., {Allen}, L.~E., {et~al.} 2004, \apjs, 154, 10,
  \dodoi{10.1086/422843}

\bibitem[{{Garhart} {et~al.}(2018){Garhart}, {Deming}, {Mandell}, {Knutson}, \&
  {Fortney}}]{GarhartEtal2018aapQatar1b}
{Garhart}, E., {Deming}, D., {Mandell}, A., {Knutson}, H., \& {Fortney}, J.~J.
  2018, \aap, 610, A55, \dodoi{10.1051/0004-6361/201731637}

\bibitem[{{Gillon} {et~al.}(2016){Gillon}, {Jehin}, {Lederer}, {Delrez}, {de
  Wit}, {Burdanov}, {Van Grootel}, {Burgasser}, {Triaud}, {Opitom}, {Demory},
  {Sahu}, {Bardalez Gagliuffi}, {Magain}, \&
  {Queloz}}]{GillonEtal2016natTRAPPIST}
{Gillon}, M., {Jehin}, E., {Lederer}, S.~M., {et~al.} 2016, \nat, 533, 221,
  \dodoi{10.1038/nature17448}

\bibitem[{{Gillon} {et~al.}(2017){Gillon}, {Triaud}, {Demory}, {Jehin}, {Agol},
  {Deck}, {Lederer}, {de Wit}, {Burdanov}, {Ingalls}, {Bolmont}, {Leconte},
  {Raymond}, {Selsis}, {Turbet}, {Barkaoui}, {Burgasser}, {Burleigh}, {Carey},
  {Chaushev}, {Copperwheat}, {Delrez}, {Fernandes}, {Holdsworth}, {Kotze}, {Van
  Grootel}, {Almleaky}, {Benkhaldoun}, {Magain}, \&
  {Queloz}}]{GillonEtal2017natTRAPPIST}
{Gillon}, M., {Triaud}, A.~H.~M.~J., {Demory}, B.-O., {et~al.} 2017, \nat, 542,
  456, \dodoi{10.1038/nature21360}

\bibitem[{{Hardy} {et~al.}(2017){Hardy}, {Harrington}, {Hardin}, {Madhusudhan},
  {Loredo}, {Challener}, {Foster}, {Cubillos}, \&
  {Blecic}}]{HardyEtal2017apjHATP13b}
{Hardy}, R.~A., {Harrington}, J., {Hardin}, M.~R., {et~al.} 2017, \apj, 836,
  143, \dodoi{10.3847/1538-4357/836/1/143}

\bibitem[{{Harris} {et~al.}(2020){Harris}, {Jarrod Millman}, {van der Walt},
  {Gommers}, {Virtanen}, {Cournapeau}, {Wieser}, {Taylor}, {Berg}, {Smith},
  {Kern}, {Picus}, {Hoyer}, {van Kerkwijk}, {Brett}, {Haldane}, {Fern{\'a}ndez
  del R{\'\i}o}, {Wiebe}, {Peterson}, {G{\'e}rard-Marchant}, {Sheppard},
  {Reddy}, {Weckesser}, {Abbasi}, {Gohlke}, \&
  {Oliphant}}]{HarrisEtal2020natNumPy}
{Harris}, C.~R., {Jarrod Millman}, K., {van der Walt}, S.~J., {et~al.} 2020,
  585, \dodoi{10.1038/s41586-020-2649-2}

\bibitem[{{Hayes} {et~al.}(2005){Hayes}, {{\"O}stlin}, {Mas-Hesse}, {Kunth},
  {Leitherer}, \& {Petrosian}}]{HayesEtal2005aaStarburst}
{Hayes}, M., {{\"O}stlin}, G., {Mas-Hesse}, J.~M., {et~al.} 2005, \aap, 438,
  71, \dodoi{10.1051/0004-6361:20052702}

\bibitem[{Hunter(2007)}]{Hunter2007cseMatplotlib}
Hunter, J.~D. 2007, Computing in Science \& Engineering, 9, 90,
  \dodoi{10.1109/MCSE.2007.55}

\bibitem[{{Ingalls} {et~al.}(2016){Ingalls}, {Krick}, {Carey}, {Stauffer},
  {Lowrance}, {Grillmair}, {Buzasi}, {Deming}, {Diamond-Lowe}, {Evans},
  {Morello}, {Stevenson}, {Wong}, {Capak}, {Glaccum}, {Laine}, {Surace}, \&
  {Storrie-Lombardi}}]{IngallsEtal2016ajSpitzerSystematics}
{Ingalls}, J.~G., {Krick}, J.~E., {Carey}, S.~J., {et~al.} 2016, \aj, 152, 44,
  \dodoi{10.3847/0004-6256/152/2/44}

\bibitem[{{Jenkins} {et~al.}(2019){Jenkins}, {Harrington}, {Challener},
  {Kurtovic}, {Ramirez}, {Pe{\~n}a}, {McIntyre}, {Himes}, {Rodr{\'{\i}}guez},
  {Anglada-Escud{\'e}}, {Dreizler}, {Ofir}, {Pe{\~n}a Rojas}, {Ribas}, {Rojo},
  {Kipping}, {Butler}, {Amado}, {Rodr{\'{\i}}guez-L{\'o}pez}, {Kempton},
  {Palle}, \& {Murgas}}]{JenkinsEtal2019mnrasProxNoTran}
{Jenkins}, J.~S., {Harrington}, J., {Challener}, R.~C., {et~al.} 2019, arXiv
  e-prints.
\newblock \doarXiv{1905.01336}

\bibitem[{{Kreidberg} {et~al.}(2018){Kreidberg}, {Line}, {Thorngren}, {Morley},
  \& {Stevenson}}]{KreidbergEtal2018apjWASP107b}
{Kreidberg}, L., {Line}, M.~R., {Thorngren}, D., {Morley}, C.~V., \&
  {Stevenson}, K.~B. 2018, \apjl, 858, L6, \dodoi{10.3847/2041-8213/aabfce}

\bibitem[{{Kreidberg} {et~al.}(2019){Kreidberg}, {Koll}, {Morley}, {Hu},
  {Schaefer}, {Deming}, {Stevenson}, {Dittmann}, {Vanderburg}, {Berardo},
  {Guo}, {Stassun}, {Crossfield}, {Charbonneau}, {Latham}, {Loeb}, {Ricker},
  {Seager}, \& {Vand erspek}}]{KreidbergEtal2019natLHS3844b}
{Kreidberg}, L., {Koll}, D. D.~B., {Morley}, C., {et~al.} 2019, \nat,
  \dodoi{10.1038/s41586-019-1497-4}

\bibitem[{{Laher} {et~al.}(2012){Laher}, {Gorjian}, {Rebull}, {Masci},
  {Fowler}, {Helou}, {Kulkarni}, \& {Law}}]{LaherEtal2012paspAPT}
{Laher}, R.~R., {Gorjian}, V., {Rebull}, L.~M., {et~al.} 2012, \pasp, 124, 737,
  \dodoi{10.1086/666883}

\bibitem[{{Lanotte} {et~al.}(2014){Lanotte}, {Gillon}, {Demory}, {Fortney},
  {Astudillo}, {Bonfils}, {Magain}, {Delfosse}, {Forveille}, {Lovis}, {Mayor},
  {Neves}, {Pepe}, {Queloz}, {Santos}, \& {Udry}}]{LanotteEtal2014aapGJ436b}
{Lanotte}, A.~A., {Gillon}, M., {Demory}, B.~O., {et~al.} 2014, \aap, 572, A73,
  \dodoi{10.1051/0004-6361/201424373}

\bibitem[{{Lee} {et~al.}(2019){Lee}, {Gommers}, {Wasilewski}, {Wohlfahrt}, \&
  {O'Leary}}]{LeeEtal2019jossPyWavelets}
{Lee}, G.~R., {Gommers}, R., {Wasilewski}, F., {Wohlfahrt}, K., \& {O'Leary},
  A. 2019, \joss, 4, 1237, \dodoi{10.21105/joss.01237}

\bibitem[{{Lewis} {et~al.}(2013){Lewis}, {Knutson}, {Showman}, {Cowan},
  {Laughlin}, {Burrows}, {Deming}, {Crepp}, {Mighell}, {Agol}, {Bakos},
  {Charbonneau}, {D{\'e}sert}, {Fischer}, {Fortney}, {Hartman}, {Hinkley},
  {Howard}, {Johnson}, {Kao}, {Langton}, \& {Marcy}}]{LewisEtal2013apjHATP2b}
{Lewis}, N.~K., {Knutson}, H.~A., {Showman}, A.~P., {et~al.} 2013, \apj, 766,
  95, \dodoi{10.1088/0004-637X/766/2/95}

\bibitem[{{Lust} {et~al.}(2014){Lust}, {Britt}, {Harrington}, {Nymeyer},
  {Stevenson}, {Ross}, {Bowman}, \& {Fraine}}]{LustEtal2014paspLeastAsym}
{Lust}, N.~B., {Britt}, D., {Harrington}, J., {et~al.} 2014, \pasp, 126, 1092,
  \dodoi{10.1086/679470}

\bibitem[{{Majeau} {et~al.}(2012){Majeau}, {Agol}, \&
  {Cowan}}]{MajeauEtal2012apjlHD189Map}
{Majeau}, C., {Agol}, E., \& {Cowan}, N.~B. 2012, \apjl, 747, L20,
  \dodoi{10.1088/2041-8205/747/2/L20}

\bibitem[{{Mansfield} {et~al.}(2020){Mansfield}, {Bean}, {Stevenson},
  {Komacek}, {Bell}, {Tan}, {Malik}, {Beatty}, {Wong}, {Cowan}, {Dang},
  {D{\'e}sert}, {Fortney}, {Gaudi}, {Keating}, {Kempton}, {Kreidberg}, {Line},
  {Parmentier}, {Stassun}, {Swain}, \& {Zellem}}]{MansfieldEtal2020apjKELT9b}
{Mansfield}, M., {Bean}, J.~L., {Stevenson}, K.~B., {et~al.} 2020, \apjl, 888,
  L15, \dodoi{10.3847/2041-8213/ab5b09}

\bibitem[{{McNamara} \&
  {O'Connell}(1992)}]{McNamaraOConnell1992apjGalColorGrad}
{McNamara}, B.~R., \& {O'Connell}, R.~W. 1992, \apj, 393, 579,
  \dodoi{10.1086/171529}

\bibitem[{{Mendon{\c{c}}a} {et~al.}(2018){Mendon{\c{c}}a}, {Malik}, {Demory},
  \& {Heng}}]{MendoncaEtal2018ajWASP43b}
{Mendon{\c{c}}a}, J.~M., {Malik}, M., {Demory}, B.-O., \& {Heng}, K. 2018, \aj,
  155, 150, \dodoi{10.3847/1538-3881/aaaebc}

\bibitem[{{Merlin} {et~al.}(2019){Merlin}, {Pilo}, {Fontana}, {Castellano},
  {Paris}, {Roscani}, {Santini}, \& {Torelli}}]{MerlinEtal2019aaAPHOT}
{Merlin}, E., {Pilo}, S., {Fontana}, A., {et~al.} 2019, \aap, 622, A169,
  \dodoi{10.1051/0004-6361/201833991}

\bibitem[{{Mighell}(2005)}]{Mighell2005mnrasPSFs}
{Mighell}, K.~J. 2005, \mnras, 361, 861,
  \dodoi{10.1111/j.1365-2966.2005.09208.x}

\bibitem[{{Miles}(2009)}]{Miles2009sassEllComet}
{Miles}, R. 2009, Society for Astronomical Sciences Annual Symposium, 28, 51

\bibitem[{{Morello}(2015)}]{Morello2015apjICA}
{Morello}, G. 2015, \apj, 808, 56, \dodoi{10.1088/0004-637X/808/1/56}

\bibitem[{Morley {et~al.}(2017)Morley, Kreidberg, Rustamkulov, Robinson, \&
  Fortney}]{MorleyEtal2017apjJWSTearth}
Morley, C.~V., Kreidberg, L., Rustamkulov, Z., Robinson, T., \& Fortney, J.~J.
  2017, The Astrophysical Journal, 850, 121, \dodoi{10.3847/1538-4357/aa927b}

\bibitem[{{Pont} {et~al.}(2006){Pont}, {Zucker}, \&
  {Queloz}}]{PontEtal2006mnrasRedNoise}
{Pont}, F., {Zucker}, S., \& {Queloz}, D. 2006, \mnras, 373, 231,
  \dodoi{10.1111/j.1365-2966.2006.11012.x}

\bibitem[{{Raftery}(1995)}]{Raftery1995BIC}
{Raftery}, A.~E. 1995, Sociological Mehodology, 25, 111

\bibitem[{{Rappaport} {et~al.}(2014){Rappaport}, {Barclay}, {DeVore}, {Rowe},
  {Sanchis-Ojeda}, \& {Still}}]{RappaportEtal2014apjKOI2700b}
{Rappaport}, S., {Barclay}, T., {DeVore}, J., {et~al.} 2014, The Astrophysical
  Journal, 784, 40, \dodoi{10.1088/0004-637X/784/1/40}

\bibitem[{{Rappaport} {et~al.}(2018){Rappaport}, {Vanderburg}, {Jacobs},
  {LaCourse}, {Jenkins}, {Kraus}, {Rizzuto}, {Latham}, {Bieryla}, \&
  {Lazarevic}}]{RappaportEtal2018mnrasExocomets}
{Rappaport}, S., {Vanderburg}, A., {Jacobs}, T., {et~al.} 2018, \mnras, 474,
  1453, \dodoi{10.1093/mnras/stx2735}

\bibitem[{{Rauer} {et~al.}(2011){Rauer}, {Gebauer}, {Paris}, {Cabrera},
  {Godolt}, {Grenfell}, {Belu}, {Selsis}, {Hedelt}, \&
  {Schreier}}]{RauerEtal2011A&Abiosig}
{Rauer}, H., {Gebauer}, S., {Paris}, P.~V., {et~al.} 2011, \aap, 529, A8,
  \dodoi{10.1051/0004-6361/201014368}

\bibitem[{{Rugheimer} {et~al.}(2015){Rugheimer}, {Kaltenegger}, {Segura},
  {Linsky}, \& {Mohanty}}]{RugheimerEtal2015ApJuvprints}
{Rugheimer}, S., {Kaltenegger}, L., {Segura}, A., {Linsky}, J., \& {Mohanty},
  S. 2015, \apj, 809, 57, \dodoi{10.1088/0004-637X/809/1/57}

\bibitem[{{Ryder}(1992)}]{Ryder1992aujphGalaxies}
{Ryder}, S. 1992, Australian Journal of Physics, 45, 395,
  \dodoi{10.1071/PH920395}

\bibitem[{{Sanchis-Ojeda} {et~al.}(2015){Sanchis-Ojeda}, {Rappaport},
  {Pall{\`e}}, {Delrez}, {DeVore}, {Gandolfi}, {Fukui}, {Ribas}, {Stassun},
  {Albrecht}, {Dai}, {Gaidos}, {Gillon}, {Hirano}, {Holman}, {Howard},
  {Isaacson}, {Jehin}, {Kuzuhara}, {Mann}, {Marcy}, {Miles-P{\'a}ez},
  {Monta{\~n}{\'e}s-Rodr{\'\i}guez}, {Murgas}, {Narita}, {Nowak}, {Onitsuka},
  {Paegert}, {Van Eylen}, {Winn}, \& {Yu}}]{SanchisOjedaEtal2015apjK2-22b}
{Sanchis-Ojeda}, R., {Rappaport}, S., {Pall{\`e}}, E., {et~al.} 2015, The
  Astrophysical Journal, 812, 112, \dodoi{10.1088/0004-637X/812/2/112}

\bibitem[{{Stevenson} {et~al.}(2012){Stevenson}, {Harrington}, {Fortney},
  {Loredo}, {Hardy}, {Nymeyer}, {Bowman}, {Cubillos}, {Bowman}, \&
  {Hardin}}]{StevensonEtal2012apjBLISShd149b}
{Stevenson}, K.~B., {Harrington}, J., {Fortney}, J.~J., {et~al.} 2012, \apj,
  754, 136, \dodoi{10.1088/0004-637X/754/2/136}

\bibitem[{{Stevenson} {et~al.}(2014){Stevenson}, {D{\'e}sert}, {Line}, {Bean},
  {Fortney}, {Showman}, {Kataria}, {Kreidberg}, {McCullough}, {Henry},
  {Charbonneau}, {Burrows}, {Seager}, {Madhusudhan}, {Williamson}, \&
  {Homeier}}]{StevensonEtal2014sciWASP43bphasecurve}
{Stevenson}, K.~B., {D{\'e}sert}, J.-M., {Line}, M.~R., {et~al.} 2014, Science,
  346, 838, \dodoi{10.1126/science.1256758}

\bibitem[{{Stevenson} {et~al.}(2017){Stevenson}, {Line}, {Bean}, {D{\'e}sert},
  {Fortney}, {Showman}, {Kataria}, {Kreidberg}, \&
  {Feng}}]{StevensonEtal2017ajWASP43bPhaseCurve}
{Stevenson}, K.~B., {Line}, M.~R., {Bean}, J.~L., {et~al.} 2017, \aj, 153, 68,
  \dodoi{10.3847/1538-3881/153/2/68}

\bibitem[{{Vanderburg} {et~al.}(2015){Vanderburg}, {Johnson}, {Rappaport},
  {Bieryla}, {Irwin}, {Lewis}, {Kipping}, {Brown}, {Dufour}, {Ciardi}, {Angus},
  {Schaefer}, {Latham}, {Charbonneau}, {Beichman}, {Eastman}, {McCrady},
  {Wittenmyer}, \& {Wright}}]{VanderburgEtal2015natDisintWD}
{Vanderburg}, A., {Johnson}, J.~A., {Rappaport}, S., {et~al.} 2015, Nature,
  526, 546, \dodoi{10.1038/nature15527}

\bibitem[{Virtanen {et~al.}(2020)Virtanen, Gommers, Oliphant, Haberland, Reddy,
  Cournapeau, Burovski, Peterson, Weckesser, Bright, {van der Walt}, Brett,
  Wilson, Millman, Mayorov, Nelson, Jones, Kern, Larson, Carey, Polat, Feng,
  Moore, {VanderPlas}, Laxalde, Perktold, Cimrman, Henriksen, Quintero, Harris,
  Archibald, Ribeiro, Pedregosa, {van Mulbregt}, \& {SciPy 1.0
  Contributors}}]{VirtanenEtal2020natmSciPy}
Virtanen, P., Gommers, R., Oliphant, T.~E., {et~al.} 2020, Nature Methods, 17,
  261, \dodoi{10.1038/s41592-019-0686-2}

\bibitem[{{Werner} {et~al.}(2004){Werner}, {Roellig}, {Low}, {Rieke}, {Rieke},
  {Hoffmann}, {Young}, {Houck}, {Brandl}, {Fazio}, {Hora}, {Gehrz}, {Helou},
  {Soifer}, {Stauffer}, {Keene}, {Eisenhardt}, {Gallagher}, {Gautier}, {Irace},
  {Lawrence}, {Simmons}, {Van Cleve}, {Jura}, {Wright}, \&
  {Cruikshank}}]{WernerEtal2004apjsSpitzer}
{Werner}, M.~W., {Roellig}, T.~L., {Low}, F.~J., {et~al.} 2004, \apjs, 154, 1,
  \dodoi{10.1086/422992}

\bibitem[{{Winn} {et~al.}(2008){Winn}, {Holman}, {Torres}, {McCullough},
  {Johns-Krull}, {Latham}, {Shporer}, {Mazeh}, {Garcia-Melendo}, {Foote},
  {Esquerdo}, \& {Everett}}]{WinnEtal2008apjXO3b}
{Winn}, J.~N., {Holman}, M.~J., {Torres}, G., {et~al.} 2008, \apj, 683, 1076,
  \dodoi{10.1086/589737}

\bibitem[{{Wong} {et~al.}(2015){Wong}, {Knutson}, {Lewis}, {Kataria},
  {Burrows}, {Fortney}, {Schwartz}, {Agol}, {Cowan}, {Deming}, {D{\'e}sert},
  {Fulton}, {Howard}, {Langton}, {Laughlin}, {Showman}, \&
  {Todorov}}]{WongEtal2015apjWASP14}
{Wong}, I., {Knutson}, H.~A., {Lewis}, N.~K., {et~al.} 2015, \apj, 811, 122,
  \dodoi{10.1088/0004-637X/811/2/122}

\end{thebibliography}

\end{document}